\documentclass[11pt]{article}

\usepackage{amsmath, amssymb, amsthm, amsfonts, latexsym, graphicx, color, hyperref}
\usepackage{epic,tikz}
\usepackage{authblk}
\usepackage{wasysym} 

\theoremstyle{plain}
\newtheorem{theorem}{Theorem}

\newtheorem{claim}{Claim}
\newtheorem{remark}{Remark}
\newtheorem{corollary}{Corollary}

\newcommand{\ii}{\mathbb{I}}

\newcommand{\norm}[1]{\left\| #1 \right\|}
\newcommand{\bra}[1]{\langle #1 \vert}

\newcommand{\ket}[1]{\vert #1 \rangle}

\newcommand{\braket}[2]{\langle #1 \vert #2 \rangle}

\newlength{\actualtopmargin}
\newlength{\actualsidemargin}
\begin{document}
\title{\textbf{Perturbative gadgets without strong interactions}}
\author[1]{Yudong Cao\thanks{cao23@purdue.edu}}
\author[2]{Daniel Nagaj\thanks{daniel.nagaj@savba.sk}}
\affil[1]{\small{Department of Computer Science, Purdue University, West Lafayette, IN 47906, USA}}
\affil[2]{\small{Institute of Physics, Slovak Academy of Sciences, D\'{u}bravsk\'{a} cesta 9, 84215 Bratislava, Slovakia}}
\date{\today}
\maketitle
\vspace{-5mm}

 
\begin{abstract}

\emph{Perturbative gadgets} are used to construct a quantum Hamiltonian whose low-energy subspace approximates a given quantum $k$-body Hamiltonian up to an absolute error $\epsilon$. Typically, gadget constructions involve terms with large interaction strengths of order $\text{poly}(\epsilon^{-1})$. Here we present a 2-body gadget construction and prove that it approximates a target many-body Hamiltonian of interaction strength $\gamma = O(1)$
up to absolute error $\epsilon\ll\gamma$ using interactions of strength $O(\epsilon)$ instead of the usual inverse polynomial in $\epsilon$. 
A key component in our proof is a new condition for the convergence of the perturbation series, allowing our gadget construction to be applied in parallel on multiple many-body terms. 
We also show how to apply this gadget construction for approximating 3- and $k$-body Hamiltonians. 

The price we pay for using much weaker interactions is a large overhead in the number of ancillary qubits, and the number of interaction terms per particle, both of which scale as $O(\text{poly}(\epsilon^{-1}))$. Our {\em strong-from-weak} gadgets have their primary application in complexity theory (QMA hardness of restricted Hamiltonians, a generalized area law counterexample, gap amplification), but could also motivate practical implementations with many weak interactions simulating a much stronger quantum many-body interaction.

\end{abstract}


\section{Introduction}

The physical properties of (quantum mechanical) spin systems can often be understood in terms of effective interactions arising from the complex interplay of microscopic interactions. Powerful methods for analyzing effective interactions have been developed, for example the renormalization group approach distills effective interactions at different length scales. Another common approach is perturbation theory -- treating some interaction terms in the Hamiltonian as a perturbation to a simple original system, giving us a sense of how the fully interacting system behaves. Here, instead of trying to understand an unknown system, we ask an engineering question: how can we build a particular (many-body) effective interaction from local terms of restricted forms? 

The idea of \emph{perturbative gadgets} provides a powerful answer to that question. Initially introduced by Kempe, Kitaev and Regev \cite{KKR06} for showing the {QMA}-hardness of {2-Local Hamiltonian} problem and subsequently used and developed further in numerous works \cite{OT06, BDOT08, BL07, JF08, BDLT08, Schuch09, CRBK14}, the perturbative gadgets are convenient tools by which arbitrary many-body effective interactions (which we call the \emph{target Hamiltonian}) can be obtained using a \emph{gadget Hamiltonian} consisting of only two-body interactions. In a broader context, these gadgets have also been used to understand the computational complexity of physical systems (e.g.\ how hard it is to determine the ground state energy) with restricted geometry of interactions \cite{OT06}, locality \cite{KKR06,OT06,BDLT08}, 
or interaction types \cite{BL07}. Here, we choose to focus on the issue of restricted coupling strengths.

In a nutshell, perturbative gadgets allow us to map between different forms of microscopic Hamiltonians. This is reminiscent of the use of gadgets in the classical theory of {NP}-completeness which starts from Constraint Satisfaction Problem ({CSP}). 
In the context of combinatorial reduction in classical complexity theory, a {\em gadget} is a finite structure which maps a set of constraints from one optimization problem into constraints of another. Using such gadgets, an instance of {3-SAT} (an {NP}-complete problem) can be efficiently mapped to an instance of Graph 3-coloring (another \textsc{NP}-complete problem \cite{GareyJohnsonNP}). On the other hand, more complex constructions allow us to create more frustrated instances of such problems without significant overhead, resulting in inapproximability as well as the existence of probabilistically checkable proofs \cite{DinurPCP07}. 

For classical {CSP} instances, gadgets can be used to reduce the arity of clauses, to reduce the size of the alphabet, or to reduce the degree of each variable on the constraint graph. It is now interesting to ask whether we could create quantum versions of gadgets that reduce the locality of interactions (analogous to arity reduction in classical CSPs), the dimension of particles (alphabet reduction) or the number of interactions that each particle is involved (degree reduction). Such gadget reductions for quantum Hamiltonians could give us tools that might help explore the way to the quantum PCP conjecture \cite{AAVqPCP13}. More modestly, gadget translations between types of local Hamiltonians would have implications  for the area law \cite{Hastings07,AALV11,ECP10} and other global properties. However, generating approximate {\em quantum} interactions from a restricted set of terms is not straightforward.

\begin{figure}
\begin{center}
\includegraphics[width=4.5cm]{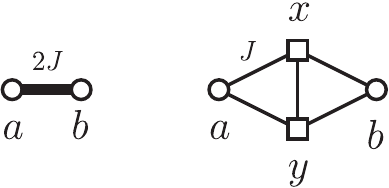}
\caption{A ferromagnetic interaction $E(a,b) = -2J a b$ of two classical spins $a,b =\in\{-1, 1\}$ can be ``built'' from half-strength interactions involving two extra ancillas. The ground states of the system on the right have $a=b$, while the lowest excited states have $a\neq b$ and energy $4J$ above the ground state energy.  Each edge between two classical spins $u$ and $v$ in this illustration represents a term $uv$ in the expression for energy. 
The $\ocircle$ nodes symbolize target spins and $\Box$ nodes are ancillas.}
\label{fig:doubleFERRO}
\end{center}
\end{figure}

Creating effective interactions with arbitrary strength by coupling a system to several ancilla degrees of freedom is a relatively simple task for classical spin systems. For example, we can create an effective (and twice stronger) ferromagnetic interaction between target spins $a,b$ using  two ancilla spins $x,y$ and connecting them to $a, b$ as illustrated in Figure~\ref{fig:doubleFERRO}. The lowest energy states of this new system correspond to the lowest energy states of a system with a ferromagnetic interaction between $a$ and $b$, with double strength. 

For general quantum interactions where the target Hamiltonian consists of many-body Pauli operators, the common {\em perturbative gadget} introduces a strongly bound ancillary system and couple the target spins to it via weaker interactions, treating the latter as a perturbation. The target many-body Hamiltonian is then generated in some low order of perturbation theory of the combined system that consists of both ancillary and target spins.
Such gadgets first appeared in the proof of {QMA}-completeness of the {2-Local Hamiltonian} problem via a reduction from {3-Local Hamiltonian} \cite{KKR06}. There they helped build effective 3-body interactions from 2-body ones.
Perturbative gadgets can also be used for reducing a target Hamiltonian with general interaction topology to a planar interaction graph \cite{OT06}, approximating certain restricted forms of 2-body interactions using other forms of 2-body interactions \cite{BL07}, realizing Hamiltonians exhibiting non-abelian anyonic excitations \cite{K14} and reducing $k$-body interactions to 2-body \cite{BDLT08,JF08}.

All existing constructions of perturbative gadgets \cite{KKR06,OT06,BDLT08,JF08,BL07,CRBK14} require interaction terms or local fields with norms much higher than the strength of the effective interaction which they generate, in order for the perturbation theory to apply\footnote{
Note that there exist special cases (e.g. Hamiltonians with all terms diagonal in the same basis) when one can analyze the Hamiltonian with non-perturbative techniques \cite{B08,OY11}.}  (see Figure \ref{fig:all2body}b). However, physically realizable systems often allow only limited interaction strengths. The main result of our paper is a way around this problem.

We first build a system with a large spectral gap between the ground state and the first excited state using many relatively weak interactions: consider a collection of $n$ spins that interact with each other via ZZ coupling of constant strength $J$. Then the first excited state of this $n$-spin system is separated from the ground energy by $\Omega(n)$. This way we can use weaker interactions to construct a \emph{core} with a large spectral gap. We then use it to replace the large local field applied onto the single ancilla (Figure \ref{fig:all2body}b) with weak interactions of a collection of ancillas (Figure \ref{fig:all2body}c). Finally, we connect the target spins to multiple ancillas instead of just one, which allows us to use weaker $\beta$ to achieve the same effective interaction strength between the target spins (Figure \ref{fig:all2body}d). 

Let us review a few definitions and then state our results precisely. An $n$-qubit\footnote{Here we use the terms \emph{qubit}, \emph{spin} and \emph{particle} interchangeably.} Hamiltonian is an $2^n \times 2^n$ Hermitian matrix; it is {\em $k$-local} (for a constant $k$) if it can be written as a sum of $M\leq \text{poly}(n)$ terms $H_j$, each acting non-trivially on at most $k$ qubits. Furthermore, we require\footnote{We use the operator norm $\|\cdot\|$, defined as $\|M\|\equiv\max_{\ket{\psi}\in\mathcal{M}}|\bra{\psi}M\ket{\psi}|$ for an operator $M$ acting on a Hilbert space $\mathcal{M}$).} $\|H_j\|\le\text{poly}(n)$, and that the entries of $H_j$ be specified by poly$(n)$ bits. The smallest eigenvalue of $H$ is its \emph{ground state energy}, and we denote it $\lambda(H)$. We use $\lambda_j(H)$ to represent the $j$-th smallest eigenvalue of $H$, hence $\lambda(H) = \lambda_1(H)$. 
Taking a 2-local Hamiltonian acting on $n$ qubits, we can associate it with an \emph{interaction graph} $G(V,E)$. Every $v\in V$ corresponds to a qubit, and there is an edge $e\in E$ between vertices $a$ and $b$ if and only if there is a non-zero 2-local term $H_e$ on qubits $a$ and $b$ such that $H_e$ is neither 1-local nor proportional to the identity operator. 
More generally, we can pair a $k$-local Hamiltonian with its {\em interaction hypergraph} in which the $k$-local terms correspond to hyper-edges involving (at most) $k$ vertices. In particular, 1-local terms correspond to self-loops. Next, because we can decompose any 2-local Hamiltonian term in the Pauli basis\footnote{
For example, the spin chain Hamiltonian $H= \sum_{i=1}^n \ket{01-10}\bra{01-10}_{i,i+1}$ has interaction edges between successive spins. Each 2-local interaction can be rewritten in the Pauli basis as $\frac{1}{4}\left(\ii\otimes \ii - X\otimes X - Y\otimes Y - Z\otimes Z\right)$. It gives us an overall energy shift (from the first term), and three Pauli edges.}, we can define a {\em Pauli edge} of an interaction graph $G$ as an edge between vertices $a$ and $b$ associated with an operator $\gamma_{ab} \, P_a\otimes Q_b$ where $P, Q \in \{\ii, X, Y, Z\}$ are Pauli matrices
 and $\gamma_{ab}$ is a real number representing the coupling coefficient. We refer to the maximum value of $|\gamma_{ab}|$ over all terms in the Hamiltonian as the \emph{interaction strength} of the Hamiltonian.
For an interaction graph where every edge is a \emph{Pauli edge}, the degree of a vertex is called its \emph{Pauli degree}. The maximum Pauli degree of vertices in an interaction graph is the Pauli degree of the graph.


\subsection{Summary}

We start in Section~\ref{sec:pert} with an introduction to the formalism of perturbation theory that we use, then 
in Section~\ref{sec:2body}, we present and prove our main result -- a construction that generates a target 2-local Hamiltonian via 
weak 2-local couplings using many ancilla particles:

\begin{theorem}[Effective 2-body interactions from weak couplings]
Let $H_\text{targ}=H_\text{else}+\sum_{j=1}^M\gamma_j \,A_{a_j}\otimes B_{b_j}$ be a 2-local Hamiltonian acting on $n$ qubits with an interaction graph of Pauli degree $p$ and ground state energy $\lambda(H_\text{targ})$, assuming $M\le\text{poly}(n)$, and the interaction strength is bounded\footnote{To deal with Hamiltonians with large interaction strengths, we can view the Pauli edges as multi-edges. We then decompose each large term into as many $O(1)$ terms as necessary.} as $\gamma_\text{max} = \max_j|\gamma_j| = O(1)$. Here $a_j,b_j\in[n]$ refers to the qubits that the $A$ and $B$ operators in the $j^\text{th}$ term act on respectively.
Let $H_\text{else}$ be some Hamiltonian with a non-negative spectrum and $\|H_\text{else}\|\le\text{poly}(n)$. 
Then for any $\epsilon>0$ and $\epsilon\ll\gamma$ there exists a 2-local Hamiltonian $\tilde{H}$ with interaction strength $O(\epsilon)$
and low-lying spectrum approximating the full spectrum of $H_{\text{targ}}$ as $|\lambda_j(\tilde{H})-\lambda_j(H_\text{targ})|\le\epsilon$ for all $j$. The gadget Hamiltonian $\tilde{H}$ acts on $n+\text{poly}(\|H_\text{else}\|,\epsilon^{-1},M)$
qubits and has an interaction graph of Pauli degree $\text{poly}(p,\|H_\text{else}\|,\epsilon^{-1},M)$.
\label{th2}
\end{theorem}

The above theorem deals with 2-local target Hamiltonians. We also know gadget constructions for reducing 3-local interactions \cite{KKR06,OT06,BDLT08} or $k$-local interactions \cite{JF08,BDLT08} to 2-local ones. These typically involve strong interaction terms, which could be further reduced to a construction consisting of only weak interactions using ideas from the proof for Theorem \ref{th2}. In particular:

\begin{theorem}[3-body terms from weak 2-body interactions]
\label{th3}
Let $H_\text{targ}=H_\text{else}+\sum_{i=1}^M\gamma_i\,A_{a_i}\otimes B_{b_i}\otimes F_{f_i}$ be a 3-local Hamiltonian acting on $n$ qubits with an interaction graph of Pauli degree $p$ and ground state energy $\lambda(H_\text{targ})$, assuming $M\le\text{poly}(n)$. Here $a_j,b_j, f_j\in[n]$ refers to the qubits that the $A$, $B$ and $F$ operators in the $j^\text{th}$ term act on respectively.
The interaction strength of $H_\text{targ}$ satisfies $\gamma_\text{max} = \max_j|\gamma_j|=O(1)$. 
Let $H_\text{else}$ be a Hamiltonian with non-negative spectrum and satisfies $\|H_\text{else}\|\le\text{poly}(n)$. 
Then for any choice $\epsilon>0$, there exists a 2-local Hamiltonian $\tilde{H}$ with interaction strength $O(\epsilon)$,
acting on $n+\text{poly}(\|H_\text{else}\|,\epsilon^{-1},M)$ qubits, with an interaction graph of Pauli degree $\text{poly}(p,\|H_\text{else}\|,\epsilon^{-1},M)$ and $|\lambda_j(\tilde{H})-\lambda_j(H_\text{targ})|\le\epsilon$ for all $j$. 
\end{theorem}
We provide basic ideas of the proof of Theorem \ref{th3} in Section \ref{sec:kbody}, hoping for a more efficient construction to be proposed in the future. 

An important property of the new construction is that it can be repeated in parallel, in essence generating arbitrary strong interaction from weak ones. Thus, we can effectively rescale interaction strengths and amplify the eigenvalue gap of a local Hamiltonian. The price we pay is the addition of many ancillas and a large increase in the number of interactions per particle.


\begin{corollary}[Coupling strength amplification by gadgets]
Let $H=\sum_{j=1}^M H_j$ be a $k$-local Hamiltonian on $n$ qubits where $M=poly(n)$ and each $H_j$ satisfies $\norm{H_j}\leq s$ for some constant $s$. 
Let $\ket{\phi_j}$ and $\lambda_j$ be the $j$-th eigenstate and eigenvalue of $H$, and let $\Delta$ be the gap between the ground and first excited energy of $H$.
Choose a magnifying factor $\theta > 1$ and an error tolerance $\epsilon>0$.
Then there exists a 2-local Hamiltonian $\tilde{H}$ with interactions of strength $O(1)$ or weaker. The low-lying eigenstates of $\tilde{H}$ are $\epsilon$-close\footnote{Here by \emph{$\epsilon$-close} we mean the norm of the difference between the two quantities (scalar, vector or matrix operator) is no greater than $\epsilon$.} to $\ket{\phi_j} \otimes \ket{0\cdots 0}_{anc}$ where $|0\cdots 0\rangle_\text{anc}$ is the state of the ancilla qubits of $\tilde{H}$, and the low-lying spectrum of $\tilde{H}$ is $\epsilon$-close to $\theta \lambda_j$. 
The low energy effective Hamiltonian of $\tilde{H}$ is $\epsilon$-close to 
$\theta H \otimes \ket{0\cdots 0}\bra{0\cdots 0}_{anc} $.
\label{thAMP}
\end{corollary}
Whereas Theorem~\ref{th2} states that a 2-local target Hamiltonian can be gadgetized to a Hamiltonian with arbitrarily weak interactions, Corollary~\ref{thAMP} states that the same could be accomplished for a $k$-local target Hamiltonian. Moreover, besides producing a gadget Hamiltonian with weak interaction that generates the target $k$-local Hamiltonian, we could also generate the target Hamiltonian multiplied by a positive factor $\theta$. In case where $\theta>1$, this can be viewed as a coupling strength amplification relative to the original target $k$-local Hamiltonian. 
The basic idea for proving Corollary \ref{thAMP} is to view the target Hamiltonian $\theta H$ (with $\theta>1$) as a sum of $O(\theta)$ copies of itself with interaction strength $O(1)$. Using the gadget constructions from \cite{JF08}, we transform the $k$-local Hamiltonian $\theta H$ to a 2-local one. 
Finally, using our 2-body gadget construction in this work, we translate this Hamiltonian to one with only weak interactions (2-body). 
What is the efficiency of this way of amplifying the couplings? We expect that the final gadget Hamiltonian acts on a system whose total number of qubits scales exponentially in $k$ (which of course is not a problem for $k=3$).


\subsection{Conclusions and Open Questions}

The gadget construction based on perturbation theory allows us to map between Hamiltonians of different types, with the same low-lying spectrum. Intuitively, the \emph{core} construction that we introduce can be thought of as replacing strong interactions by {\em repetition} of interactions with ``classical'' ancillas; it works because for a low-energy state, all our extra qubits are close to the state $\ket{0}$. This is reminiscent of repetition encoding found e.g. in \cite{YoungLidar}. Another feature of 2-body gadget that we use heavily is {\em parallelization}; it is crucial to show that the perturbation series converges even with many gadgets, relaxing the usual assumption about the norm of the perturbation.

The 2-body gadget constructed here should find use in computer science as well as physics.
First, in complexity theory, Theorem~\ref{th2} together with \cite{KKR06} or Theorem~\ref{th3} with \cite{New3local} implies QMA-completeness of the 2-local Hamiltonian problem with $O(1)$ terms and an $O(1)$ promise gap.
As a consequence, we also obtain efficient universality for quantum computation with time-independent, 2-local Hamiltonians with restricted form/strength of terms, complementing \cite{JanzingWocjanErgodic, OT06, Universal2local}.
	Second, our amplification method from Corollary~\ref{thAMP} has been utilized in a counterexample to the generalized area law in \cite{AALNSV14}.
	Finally, we envision practical experimental applications of Theorem~\ref{th2} -- strengthening effective interactions between target (atomic) spins through many coupled mediator spins.

Thinking further about interaction strengths and spectra of local Hamiltonians, we realize that 
Corollary~\ref{thAMP} allows us to amplify the {\em eigenvalue gap} (low eigenvalue spacing) of a Hamiltonian.
Does it have direct implications for hardness of Local Hamiltonian problems?
When we use it on Hamiltonians appearing in QMA-complete constructions, the fractional {\em promise gap} (the ratio of the number of frustrated terms to the number of all terms in the Hamiltonian for a ground state of a local Hamiltonian) gets smaller. Thus, it does not provide us with a tool that would directly help us to move towards the quantum PCP conjecture \cite{AAVqPCP13}. Nevertheless, we have added another tool for mapping between Hamiltonians to our repertoire.

An important problem remains open. The price we pay for our construction is a massive blowup in the degree (the number of interactions per particle). Is there a possibility of a quantum degree-reduction gadget? One might try to use a ``bad'' quantum code for encoding each spin into several particles, whose encoded low-weight operators that can be implemented in {many} possible ways; this does not seem possible for both $X$ and $Z$ operators. 
As things stand, without a degree-reduction gadget, we do not have a way to reproduce our results in simpler geometry. It would be really interesting if one indeed could create $O(1)$-norm effective interactions from $O(1)$-terms in 3D or even 2D lattices.

We also need to think about the robustness of our results -- what will change when the Hamiltonians are not exactly what we asked for? How precise do we need to be {\em e.g.\ }for the 3-body to 2-body gadgets, so that the second- and first-order terms get canceled?
Next, we realize that given an absolute error $\epsilon$, we do not have an algorithm that computes the optimal assignments to $R, C, \beta,$ {\em etc.\ }such that the resulting error is close to but smaller than $\epsilon$ (as in the case of \cite{CRBK14}). 
Finally, Bravyi, Terhal, DiVincenzo and Loss \cite{BDLT08} mentioned whether a $k$-body to 2-body reduction can be implemented with poly$(k)$ overhead in interaction strength (instead of exponential in $k$). This question remains open. In \cite{BDLT08} the exponential scaling in the overhead is due to the usual gadget constructions which require poly$(\epsilon^{-1})$ interaction strength. We hope that with our new gadget construction, the results in \cite{BDLT08} could be improved.


\section{Effective interactions based on perturbation theory}
\label{sec:pert}

The purpose of a perturbative gadget is to approximate a target $n$-qubit Hamiltonian $H_\text{targ}$ by a gadget Hamiltonian $\tilde{H}$ which uses a restricted form of interactions among the $n$ qubits that $H_\text{targ}$ acts on and poly($n$) additional ancilla qubits. The subspace spanned by the lowest $2^n$ eigenstates of $\tilde{H}$ should approximate the spectrum of $H_\text{targ}$ up to a prescribed error tolerance $\epsilon$ in the sense that the $j^\text{th}$ lowest eigenvalue of $\tilde{H}$ differs from that of $H_\text{targ}$ by at most $\epsilon$ and the inner product between the corresponding eigenstates of $\tilde{H}$ and $H_\text{targ}$ (assume no degeneracy) is at least $1-\epsilon$. These error bounds can be established using perturbation theory \cite{KKR06,OT06}. There are various versions of perturbation theory available for constructing and analyzing gadgets (for a review see \cite{BDL11}). For example, Jordan and Farhi \cite{JF08} use Bloch's formalism, while Bravyi et al.\ rely on the Schrieffer-Wolff transformation \cite{BDLT08}. For our gadget construction in Section \ref{sec:2body}, we use the technique from \cite{KKR06,OT06}. 

Let us now review the basic ideas underlying the construction of effective Hamiltonians from gadgets.
The gadget Hamiltonian $\tilde{H}=H+V$ is a sum of an unperturbed Hamiltonian $H$ and a perturbation $V$. $H$ acts only on the ancilla space, energetically penalizing certain configurations, and favoring a specific ancilla state or subspace. Second, we have a perturbation $V$ describing how the target spins interact with the ancillas. 

Let us introduce the following notations: let $\lambda_j$ and $|\psi_j\rangle$ be the $j^\text{th}$ eigenvalue and eigenvector of $H$ and similarly define $\tilde\lambda_j$ and $|\tilde\psi_j\rangle$ for $\tilde{H}$, assuming all the eigenvalues are labeled in a weakly increasing order ($\lambda_1\le\lambda_2\le\cdots$, similarly for $\tilde{\lambda}_j$). Using a cutoff value $\lambda_*$, let us call $\mathcal{L}_-=\text{span}\{\ket{\psi_j}\,|\,\lambda_j\le\lambda_*\}$ the {\em low-energy subspace} and $\mathcal{L}_+=\text{span}\{\ket{\psi_j}\,|\,\lambda_j>\lambda_*\}$ the {\em high-energy subspace}. Let ${\Pi_-}$ and ${\Pi_+}$ be the orthogonal projectors onto the subspaces $\mathcal{L}_-$ and $\mathcal{L}_+$. For an operator $O$ we define the partitioning of $O$ into these subspaces as $O_-={\Pi_-}O{\Pi_-}$, $O_+={\Pi_+}O{\Pi_+}$, $O_{-+}={\Pi_-}O{\Pi_+}$ and $O_{+-}={\Pi_+}O{\Pi_-}$. We define similar notations $\tilde{\mathcal{L}}_-$ and $\tilde{\mathcal{L}}_+$ for $\tilde{H}$.

Our first goal is to understand $\tilde{H}_-$, the restriction of the gadget Hamiltonian to the low-energy subspace. Let us consider the operator-valued {\em resolvent} $\tilde{G}(z)=(z\ii-\tilde H)^{-1}$ where $\ii$ is the identity operator. Similarly let us define $G(z)=(z\ii -H)^{-1}$. Note that $\tilde{G}^{-1}(z)-G^{-1}(z)=-V$, which allows an expansion of $\tilde{G}$ in powers of $V$: 
\begin{equation}\label{eq:G_expand}
\tilde{G}=(G^{-1}-V)^{-1}=G(\ii-VG)^{-1}=G+GVG+GVGVG+\cdots.
\end{equation}
It is also standard to define the {\em self-energy} $\Sigma_-(z)=z\ii-({\tilde G}_-(z))^{-1}$. It is important because the spectrum of $\Sigma_-(z)$ gives an approximation to the spectrum of $\tilde{H}_-$, since by definition $\tilde{H}_-=z\ii-{\Pi_-}(\tilde{G}^{-1}(z)){\Pi_-}$ while $\Sigma_-(z)=z\ii-({\Pi_-}\tilde G(z){\Pi_-})^{-1}$. As explained in \cite{OT06}, if $\Sigma_-(z)$ is roughly constant in some range of $z$ (see Theorem \ref{th:perturbation} below for details) then $\Sigma_-(z)$ is (loosely speaking) playing the role of $\tilde{H}_-$. This was formalized in Theorem 3 in \cite{KKR06} (and improved in Theorem A.1 in \cite{OT06}). Similarly to \cite{OT06}, we choose to work with $H$ whose lowest eigenvalue is zero and whose spectral gap is $\Delta$. However, for establishing our main results as stated in Theorems~\ref{th2} and \ref{th3}, here we use a slightly modified version of Theorem 3 from \cite{KKR06}:

\begin{theorem}[Gadget approximation theorem, modified from \cite{KKR06}]
\label{th:perturbation}
Let $H$ be a Hamiltonian with a gap $\Delta$ between its ground state and first excited state. Assuming the ground state energy of $H$ is 0, let $\lambda_*=\Delta/2$. Consider a bounded norm perturbation $V$. The perturbed Hamiltonian is then $\tilde{H}=H+V$. Following the notations introduced previously, if the following holds:
\begin{enumerate}
\item $\tilde{\mathcal{L}}_-\cap \mathcal{L}_+=\varnothing$, \emph{i.e.\ }when the perturbed low- and the unperturbed high-energy states are separated: the energy $\bra{\xi_+}{H}\ket{\xi_+} > \lambda_* > \bra{\xi_-}\tilde{H}\ket{\xi_-}$ for any normalized state $\ket{\xi_+} \in \mathcal{L}_{+}$ and $\ket{\xi_-} \in \tilde{\mathcal{L}}_{-}$.

\item There is an effective Hamiltonian $H_\text{eff}$ with a spectrum contained in $[E_1,E_2]$ for some $\epsilon>0$ and $E_1<E_2<\Delta/2-\epsilon$, such that for every $z\in[E_1-\epsilon,E_2+\epsilon]$, 
the self-energy $\Sigma_-(z)$ obeys
$\|\Sigma_-(z)-H_\text{eff}\|\le\epsilon$.
\end{enumerate}
then all the eigenvalues of $\tilde{H}_-$ are close to the eigenvalues of $H$, obeying 
\[|\lambda_j(H_\text{eff})-\lambda_j(\tilde{H}_{-})|\le\epsilon.\]

\end{theorem}

Here is how our statement of Theorem \ref{th:perturbation} differs from the original version \cite{KKR06}. There, $\|V\|\le\Delta/2$ is used as a condition, while we only assume that $\tilde{\mathcal{L}}_-\cap \mathcal{L}_+=\varnothing$, meaning that the energy levels of the perturbed low-energy subspace do not mix with the unperturbed excited subspace. Our assumption is weaker, yet it is sufficient to yield the same conclusion as Theorem 3 in \cite{KKR06}. We choose to avoid enforcing $\|V\|\le\Delta/2$ in our construction since it imposes limitations on the \emph{global} properties of the gadget construction, in particular the number of ancillas we use, disregarding the structure of the perturbation. 
One might question whether the use of perturbation theory is sensible if we assume that the perturbation $\|V\|$ is no longer small compared to the spectral gap $\Delta$. We note that such use has been justified previously by Bravyi et al. \cite{BDLT08} in a similar context. 

To apply Theorem \ref{th:perturbation}, a series expansion for the self-energy $\Sigma_-(z)=z\ii-\tilde{G}_-^{-1}(z)$ is truncated at some low order, for which $H_\text{eff}$ is approximated. 
Using the series expansion of $\tilde{G}$ in \eqref{eq:G_expand}, the self-energy can be expanded as (see \cite{KKR06} for details)
\begin{equation}\label{eq:selfenergy}
	\Sigma_-(z) = H_-+V_-+V_{-+}G_+(z)V_{+-}+V_{-+}G_+(z)V_+G_+(z)V_{+-}+\cdots,
\end{equation}
with $G_+(z)=\Pi_+ (z\ii -H)^{-1} \Pi_+$. 
The $2^\text{nd}$ and higher order terms in this expansion give rise to effective many-body interactions.
Introducing auxiliary spins and a suitable selection of 2-local $H$ and $V$, we can engineer $\Sigma_-(z)$ to be $\epsilon$-close to $H_\text{eff}=H_\text{targ}\otimes\Pi_-$ (here $\Pi_-$ is the projector to the ground state subspace of the ancillas) in the range of $z$ considered in Theorem~\ref{th:perturbation}. 
Therefore with $\|\Sigma_-(z)-H_\text{eff}\|\le\epsilon$, condition 2 of Theorem~\ref{th:perturbation} is satisfied.

In the next Section, we will look at the usual 2-body gadgets and see how the second order terms in the self-energy result in the desired effective Hamiltonian. Then we present our construction that involves more ancillas with weaker interactions, and show that the effective Hamiltonian is again what we want, and both conditions for Theorem 3 are satisfied.

\section{A new gadget for 2-body interactions}\label{sec:2body}

We can decompose any 2-local interaction of spin-$\frac{1}{2}$ particles into the Pauli basis\footnote{It is useful that the Pauli matrices $A,B \in \{\ii,X,Y,Z\}$ square to identity, because $A^2$ and $B^2$ terms in our effective Hamiltonian will become simple overall energy shifts}, 
using terms of the form $\gamma A\otimes B$, with the operator $A$ acting on spin $a$ and $B$ acting on spin $b$, and $\gamma$ the {interaction strength}. 
It will be enough to show how to replace any such Pauli' interaction in our system by a gadget, aiming at the target interaction 
$H_\text{targ}=H_\text{else}+\gamma A\otimes B$, with $H_\text{else}$ some $O(1)$-norm, 2-local Hamiltonian.
First, we briefly review the existing constructions \cite{OT06,BDLT08,CRBK14} for generating $H_\text{targ}$ using a gadget Hamiltonian $\tilde{H}$. Then we present a new 2-body gadget which simulates an arbitrary $\gamma = O(1)$ strength 2-local interaction using a gadget Hamiltonian with terms of strength only $o(1)$, ``building'' quantum interactions from many weaker ones.
$\quad$\\

\begin{figure}
\begin{center}
\includegraphics[width=15.5cm]{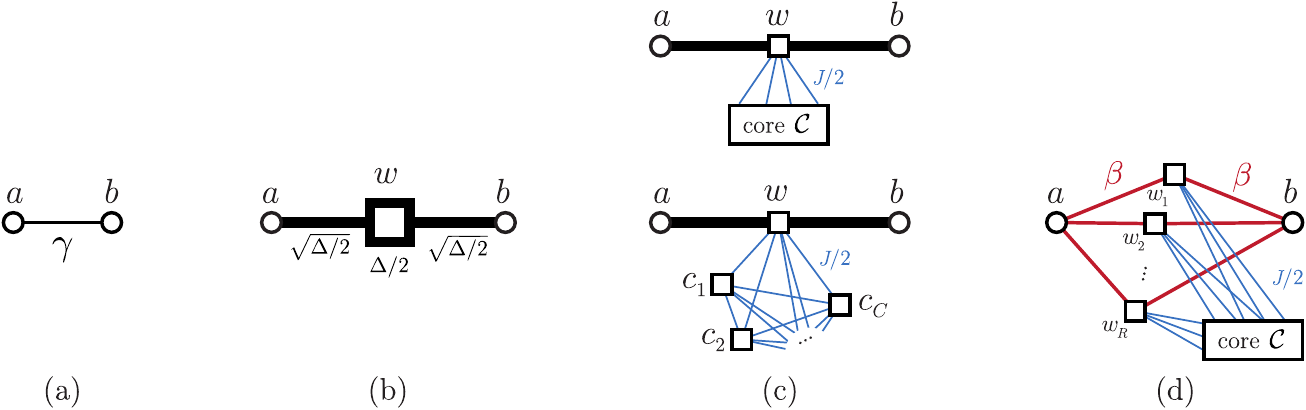}
\caption{Effective two-body interaction mediated by ancilla qubits. Each node represents a particle. The size of the node indicates the strength of local field applied onto it. 
The width of each edge shows the strength of the interaction between the particles that the edge connects.
(a) The desired 2-local interaction between target spins $a,b$.
(b) The usual perturbative gadget uses a single ancilla $w$ in a strong local field, and large-norm interactions with the target spins.
(c) We can replace the strong local field $\Delta/2$ by ferromagnetic interactions with a fixed {\em core} -- a group of $C$ ``core'' ancilla qubits located in a field of strength $J/2$, interacting with each other ferromagnetically (as a complete graph), with strength $J/2$.
(d) Instead of the strong interactions between target spins $a,b$ and a single ancilla $w$, we can use $R$ different ``direct'' ancillas (labelled as $w_1$, $w_2$, $\cdots$, $w_R$) and weaker interactions of strength $\beta$.} 
\label{fig:all2body}
\end{center}
\end{figure}

\noindent{\emph{The usual construction.}} Consider a target 2-local term involving two qubits $a,b$ as depicted in Figure~\ref{fig:all2body}(a). The standard construction of a gadget Hamiltonian $\tilde{H}$ that captures the 2-local target term is shown in Figure \ref{fig:all2body}(b). First, we introduce an ancilla qubit $w$ bound by a local field, with the Hamiltonian $H = - \frac{\Delta}{2} Z_w$. Alternatively, up to a spectral shift we could write $H=\Delta|1\rangle\langle{1}|_w$ where $|1\rangle\langle{1}|_w=\frac{1}{2}(\ii-Z_w)$. 
Then we let $w$ interact with $a$ and $b$ through $\sqrt{\Delta/2}\, A \otimes \ii \otimes X_w$ and $-\sqrt{\Delta/2}\, \ii \otimes B \otimes X_w$, 
and choose $\Delta = \Theta(\epsilon^{-1})$, according to the analysis in \cite{CRBK14}.
We can view these terms as a perturbation to $H$ and show that up to an error $\epsilon$, the low energy effective Hamiltonian as calculated via Equation \ref{eq:selfenergy} is approximately $A\otimes B \otimes|0\rangle\langle{0}|_w$ (up to an overall energy shift)  \cite{CRBK14}.  
Here ``up to an error $\epsilon$" means that the $j$-th lowest eigenvalue of $\tilde{H}$ differs from that of $H_\text{targ}$ by at most $\epsilon$ and the inner product between the corresponding eigenstates of $\tilde{H}$ and $H_\text{targ}$ (assume no degeneracy) is at least $1-\epsilon$.
$\quad$\\

\noindent{\emph{Our construction.}} In the usual construction, with better precision (decreasing $\epsilon$), the spectral gap $\Delta$ (related to local field strength) and interaction strengths grow as \emph{inverse} polynomials in $\epsilon$. We now suggest a 2-body gadget which simulates an arbitrary $O(1)$ strength target interaction using a gadget Hamiltonian of only $O(\epsilon)$ interaction strength \emph{i.e.\ }without the need for large-norm terms. We build it in a sequence of steps illustrated in Figure \ref{fig:all2body}.


The first step is to reduce the large local field $\Delta$ in Figure~\ref{fig:all2body}(b).
Let us call the ancilla $w$ directly interacting with the target spins a {\em direct} ancilla.
We add a {\em core} $\mathcal{C}$ -- a set of $C$ ancilla qubits $c_1,\dots,c_C$, with a complete graph of ferromagnetic (ZZ) interactions of strength $\frac{J}{2}$, and in a local field of strength $\frac{J}{2}$ where $J=O(\epsilon)$.
We then let the direct ancilla $w$ interact (ferromagnetically) with each of the core ancillas, as in Figure~\ref{fig:all2body}(c).
The Hamiltonian for the direct and core ancillas then reads
\begin{align}
	{\frac{J}{2}\sum_{c\in \mathcal{C}}(\ii-Z_w Z_c)}
	+ \underbrace{\frac{J}{2}\sum_{c\in \mathcal{C}}(\ii-Z_c) 
	+ \frac{J}{2}\sum_{c,c' \in \mathcal{C}}(\ii-Z_c Z_{c'})}_{\equiv H_\mathcal{C}}.
	\label{H2bodySingleCore}
\end{align}
$H_\mathcal{C}$ is the Hamiltonian describing the core $\mathcal{C}$.
The ground state of this Hamiltonian is $\ket{0}_w \ket{0\cdots 0}_{\mathcal{C}}$, and the gap between its ground and first excited state $\ket{1}_w \ket{0\cdots 0}_{\mathcal{C}}$ is $\Delta = J C$. Here $C$ is the number of ancillas in the core $\mathcal{C}$.

The second step is to use $R$ direct ancillas $w_1,\dots,w_R$ instead of just one, connecting each of them to the core ancillas as in Figure~\ref{fig:all2body}(d). The Hamiltonian then becomes
\begin{align}
	H &= \frac{J}{2} \sum_{i=1}^{R} \sum_{c\in \mathcal{C}}(\ii-Z_{w_i} Z_c) + H_\mathcal{C}.
\label{H2bodyTCore} 
\end{align}
Its ground state is $\ket{0\cdots 0}_w \otimes \ket{0\cdots 0}_{\mathcal{C}}$ (here we use the subscript $w$ to refer to all the direct ancillas connected to the target qubits), and the gap between the two lowest energies is still $\Delta =JC$.

We want to engineer an effective interaction $H_\text{targ}=\gamma A_a\otimes B_b + H_\text{else}$, where  
the first term is our desired Hamiltonian, and $H_\text{else}$ is a finite-norm Hamiltonian that includes all the other terms that we want to leave unchanged by this gadget. Starting with the Hamiltonian $H$ in \eqref{H2bodyTCore}, we add a perturbation
\begin{align}
	\label{eq:V}
	V = H_\text{else}
	 + \beta \sum_{i=1}^{R} \left(A_a \otimes X_{w_i} - B_b\otimes X_{w_i}\right),
\end{align}
where $\beta>0$ is the strength of the interactions between the target spins and the direct ancillas.
Showing that we can use perturbation theory to obtain the effective Hamiltonian crucially relies on Theorem~\ref{th:perturbation}, and we will justify that its conditions hold later. Let us now prepare the notations and tools for this.
Let $\mathcal{L}_-$ be the subspace with the ancillas in the state $\ket{0}^{\otimes (R+C)}$, let the subspace of all unit vectors in the Hilbert space that is orthogonal to $\mathcal{L}_-$ be denoted as $\mathcal{L}_+$, and let $\Pi_-$ and $\Pi_+$ be the projectors onto these subspaces. We then have  
\begin{align}
	V_- &= \Pi_-V\Pi_-=H_\text{else}\otimes\Pi_-, \\
	V_{-+} &= \Pi_-V\Pi_+=\beta(A_a-B_b)\otimes \sum_{i=1}^{R} \ket{0}\langle{1}|_{w_i}, \\
	V_{+-} &= \Pi_+V\Pi_-=\beta(A_a-B_b)\otimes \sum_{i=1}^{R}\ket{1}\langle{0}|_{w_i},	 \\
	V_+ &= \Pi_+V\Pi_+ = H_\text{else}\otimes\Pi_+
	  + \beta \sum_{i=1}^R (A_a-B_b)\otimes \Pi_+X_{w_i} \Pi_+.
\end{align}
The self-energy expansion which describes the low-energy sector of the gadget Hamiltonian $\tilde{H}=H+V$ becomes\footnote{For details see Section \ref{sec:pert}.}
\begin{align}\label{eq:sigma_z}
	\Sigma_-(z)
	=\underbrace{H_\text{else}}_\text{$1^\text{st}$ order}
	  +\underbrace{\frac{1}{z-\Delta}R\beta^2(A_a-B_b)^2}_\text{$2^\text{nd}$ order}
	  +\underbrace{\sum_{m=1}^\infty V_{-+}G_+(V_+G_+)^m V_{+-}}_\text{error term}.
\end{align}
Recall that $G(z)=(z\ii-H)^{-1}$. The range of $z$ we consider is $|z| \leq \|H_\text{else}\|+|\gamma|$. Assuming\footnote{For negative $\gamma$, we replace the factor $(A_a - B_b)$ in \eqref{eq:V} with $(A_a-\text{sgn}(\gamma)B_b)$.} 
$\gamma>0$ in $H_\text{targ}$, we can choose 
\begin{align}
	\beta = \sqrt{\frac{\gamma\Delta}{2R}} = \sqrt{\frac{\gamma JC}{2R}}.
	\label{betachoice}
\end{align}
Since $z\ll\Delta$, we can write $\frac{1}{z-\Delta}=-\frac{1}{\Delta}\left(1-\frac{z}{\Delta}\right)^{-1}\approx -\frac{1}{\Delta}+O\left(\frac{1}{\Delta^2}\right)$. Then the $1^\text{st}$ and $2^\text{nd}$ order terms are approximately equal to the desired effective Hamiltonian $H_\text{eff}=H_\text{targ}\otimes\Pi_-$ up to an overall spectral shift (because $A^2 = B^2 = \ii$).
We will show later, in Claim \ref{claim:perturb}, that with good choices of $R$ and $C$ we can 
make $\beta$ and $J$ as small as we want.
$\quad$\\

\begin{figure}
\begin{center}
\includegraphics[width=6cm]{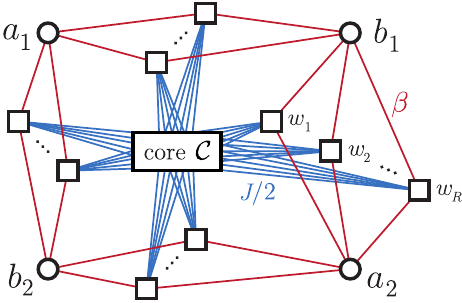}
\caption{Parallel composition of $M$ (here $M=4$) two-body gadgets from Figure~\ref{fig:all2body}(d), using a single common core with $C$ ``core'' ancillas. Each gadget has $R$ ``direct'' ancillas interacting with the target spins.  The total number of ancillas is thus $MR+C$.}
\label{fig:singlecore}
\end{center}
\end{figure}

\noindent{\emph{Parallel 2-body gadgets.}} So far, we have focused on a single 2-local term in our target Hamiltonian (see Figure \ref{fig:all2body}). Similarly to \cite{OT06}, we can apply our gadgets \emph{in parallel}, which enables us to deal with a target Hamiltonian with $M$ such 2-local terms. 
Let us then consider a target Hamiltonian of the form 
\begin{equation}\label{eq:2targ}
H_\text{targ}=H_\text{else}+\sum_{j=1}^M\gamma_j A_{a_j}\otimes B_{b_j} 
\end{equation}
and apply our construction to every term $\gamma_j A_{a_j}\otimes B_{b_j}$ in parallel, as in Figure~\ref{fig:singlecore}.
Note that we save a lot of resources by using only a single core. Each target term $\gamma_j A_{a_j}\otimes B_{b_j}$ is associated with $R$ direct ancilla qubits $w_1^{(i)}$, $w_2^{(i)}$, $\cdots$, $w_R^{(i)}$ that are connected to target spins $a_i$ and $b_i$. All of the direct ancillas also interact with each of the $C$ core ancillas. As before, the core consists of $C$ qubits that are fully connected with ferromagnetic (ZZ) interactions of strength $\frac{J}{2}$ and also with local fields of strength $\frac{J}{2}$ on each qubit. 
Hence the full gadget Hamiltonian for the general 2-local target Hamiltonian in \eqref{eq:2targ} takes the form $\tilde{H}=H+V$ with
\begin{align}
	H & = \frac{J}{2} \sum_{j=1}^{M}\sum_{i=1}^{R} \sum_{c\in \mathcal{C}}(\ii-Z_{w_i^{(j)}} Z_c)+ H_\mathcal{C}, \label{2localgadget_full}\\
	V & =   H_\text{else}+\sum_{j=1}^M \beta_j \sum_{i=1}^{R} \left(A_{a_j}  - B_{b_j}\right)\otimes X_{w_i^{(j)}}. \nonumber
\end{align}
where $H_\mathcal{C}$ is the core Hamiltonian from \eqref{H2bodyTCore}, 
$\beta_j=\sqrt{\frac{\gamma_jJC}{2R}}$ and the spectral gap between the ground state and the first excited state of $H$ is $\Delta=JC$. Computing the self-energy expansion as in \eqref{eq:sigma_z} for the gadget Hamiltonian in \eqref{2localgadget_full} yields a contribution $-\frac{1}{z-\Delta}\sum_{j=1}^M\beta_j^2R(A_{a_j}-B_{b_j})^2$ at the second order (see Claim \ref{claim:perturb} for more details). Because each term in the perturbative expansion $\Sigma_-(z)$ corresponds to a sequence of state transitions from $\mathcal{L}_-$ to $\mathcal{L}_+$ and back\footnote{Note that in fact $\mathcal{L}_-=\text{span}\{|0\cdots 0\rangle_w|0\cdots 0\rangle_\mathcal{C}\}$ where the subscript $w$ refers to all the ancillas $w_1^{(j)}$, $w_2^{(j)}$, $\cdots$, $w_R^{(j)}$, for $j=1,2,\cdots,M$. The transitions that contribute to the perturbative expansion $\Sigma_-(z)$ are restricted to only to the direct ancillas $|0\cdots 0\rangle_w$ since the core ancillas do not interact with the target qubits.}, the second order contribution comes from those transitions where one ancilla is flipped from $|0\rangle$ to $|1\rangle$ and back to $|0\rangle$. Such transitions cannot involve more than one ancilla qubit. Hence we can regard the second order transitions involving different ancillas as occuring independently of each other, or \emph{in parallel}. This ``parallelism" enables the 2-body gadgets to capture multiple 2-local target terms. Let us now clarify the notions of \emph{parallel} and \emph{serial} application of gadgets.

\begin{remark}(Parallel gadgets) Given a target Hamiltonian $H_\text{targ}=H_\text{else}+\sum_{j=1}^M H_\text{targ,$j$}$ that is a sum of multiple components $H_\text{targ,$j$}$, parallel application of gadgets combines individual gadget constructions for each $H_\text{targ,$i$}$ to form a gadget Hamiltonian $\tilde{H}$ such that its low-lying spectrum captures that of $H_\text{targ}$.
\end{remark}

\begin{remark}(Serial gadgets)
Suppose some target Hamiltonian $H_\text{targ}$ is approximated by a gadget Hamiltonian $\tilde{H}^{(1)}$ such that its low-lying sector approximates the spectrum of $H_\text{targ}$. If one further replaces some terms in $\tilde{H}^{(1)}$, creating another gadget Hamiltonian $\tilde{H}^{(2)}$ whose low-lying spectrum captures the spectrum of $\tilde{H}^{(1)}$, we say that the two gadgets are applied in series to reduce $H_\text{targ}$ to $\tilde{H}^{(2)}$.
\end{remark}

In order to show that the low-lying subspace of our gadget Hamiltonian $\tilde{H}$ captures the spectrum of $H_\text{targ}$ using Theorem~\ref{th:perturbation}, it is necessary to establish that $\tilde{H}$ meets both conditions of the theorem. The first condition ($\tilde{\mathcal{L}}_-\cap\mathcal{L}_+=\varnothing$) requires the perturbed low-energy subspace of the gadget Hamiltonian to be disjoint from the unperturbed high-energy subspace. We will prove this as Claim \ref{Lpm} below. The second condition says that the self-energy expansion $\Sigma_-(z)$ can be approximated by an effective Hamiltonian when $z$ is in a certain range. We establish this as Claim~\ref{claim:perturb} for $\tilde{H}$ by proving that the perturbation series converges for $\Sigma_-(z)$.  Theorem \ref{th2} then follows from Theorem~\ref{th:perturbation} with $\tilde{H}$ being the Hamiltonian in \eqref{2localgadget_full}.


\subsection{The 2-local construction satisfies the subspace condition.}
\label{sec:subspaces2body}

The first condition in Theorem~\ref{th:perturbation} is a property of the high-energy subspace of the original Hamiltonian. We need it in order to avoid the need to bound the norm of the whole perturbation. Let us provide a high-level description of the condition and the ideas behind its proof.

\begin{figure}
\begin{center}
\includegraphics[width=12cm]{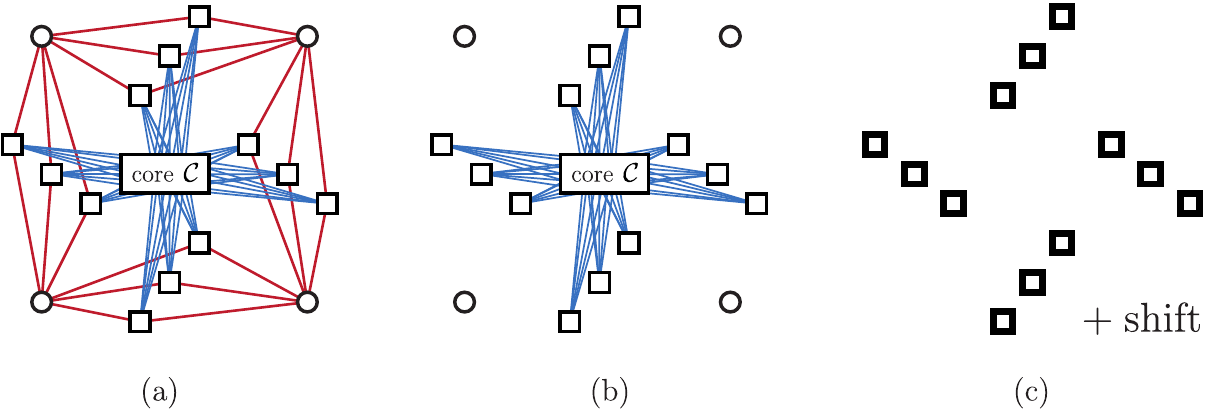}
\caption{A sequence of gadget Hamiltonians with progressively lower bounds on $E_+$. (a) Taking the terms acting on the target spins to be all the same. (b) Decoupling the target spins from the direct ancillas using $-\ii$ operators on the target spins and (weighted) $X$-fields on the direct ancillas.
(c) Replacing the interactions with core ancillas by an overall shift, and a (weighted) $Z$-field on the direct ancillas, arriving at \eqref{eq:1local2}.}
\label{fig:simplifyH}
\end{center}
\end{figure}

Consider the gadget Hamiltonian $\tilde{H} = H + V$ defined in \eqref{2localgadget_full}.
We need to lower bound the lowest energy $E_+ = \min_{\psi} \bra{\psi}\tilde{H}\ket{\psi}$ of a state $\ket{\psi}$ that comes from the subspace $\mathcal{L}_+$, the excited subspace of $H$, spanned by states orthogonal to the state $\ket{0\cdots 0}_w$. The terms in $H$ involve only ancilla qubits, while $V$ includes $H_{\text{else}}$, and terms that couple some computational (target) qubit $a$ and a direct ancilla $w$. These 2-local terms have form $\beta_{w}A_{a} \otimes X_{w}$, with 
interaction strengths $\left|\beta_w\right| \leq \beta_\text{max} = O(1)$, as in Figure~\ref{fig:simplifyH}(a).
We now want to show that $E_+$ is strictly above $\lambda_*= \frac{\Delta}{2}$.
To do this, we lower bound $E_+$ using a sequence of progressively simpler Hamiltonians, ending up with 1-local ones in \eqref{eq:1local}, \eqref{eq:Hamax} and \eqref{eq:1local2}. 

Consider the following sequence of lower bounds
(see Appendix~\ref{sec:EplusSimplification} for a detailed derivation).
First, $E_+$ for the general Hamiltonian $\tilde{H}$ is greater or equal to the value of $E_+$ for a similar system in which all of the operators $A_a$ are the same (so that they do not compete against each other in lowering the energy) as in Figure~\ref{fig:simplifyH}(a). Second, we can only lower $E_+$ by making all of the operators $A_{a}$ identities, 
and using only operators $- \left|\beta_w\right| X_w$ on the direct ancillas. 
Because the target spins are now independent from the ancillas, the contribution from $H_{\text{else}}$ is then no larger in magnitude than $\norm{H_{\text{else}}}$.
This is depicted in Figure~\ref{fig:simplifyH}(b). 

We are now left with a Hamiltonian which is a sum of $H_{\text{else}}$, single-qubit terms on the direct ancilla qubits, and their interactions\footnote{If the values of $\beta$ are different for different bonds, we still use a single core with a fixed $J$, fixed $C$, fixed $\Delta = JC$,  and adjust each $\beta_w$ for each target interaction individually so that the resulting effective interaction strength $\frac{\beta_w^2 R}{\Delta} = O(1)$ is what we desire.} with the core ancillas.
The Hilbert space divides into a direct sum of invariant subspaces labeled by the state of the core ancillas. These subspaces are decoupled (the original Hamiltonian $H$ and the perturbation $V$ do not flip the core ancillas), so we can analyze them one by one.
We do so for the subspaces with $a\geq 1$ core ancillas flipped to $\ket{1}$, and then finally for the subspace with all core ancillas equal to $\ket{0}$.
It turns out that in each such subspace we can map the terms $Z_w Z_c, Z_c Z_{c'}, Z_c$ and $H_{\text{else}}$ of the Hamiltonian\footnote{Note that there is no direct $Z_w$ term on the direct ancillas, so that a single direct ancilla flip increases the energy by $\Delta=JC$.}
 to one that is simply an overall shift, and a $-\frac{\Delta_a}{2} Z_w$ term on each of the direct ancillas, with $\Delta_a$ a function of how many ancillas were flipped. 
The resulting 1-local Hamiltonian illustrated in Figure~\ref{fig:simplifyH}(c) can be analyzed, and yields the desired lower bound on $E_+$. Let us then state and prove our first Claim.

\begin{claim}\label{Lpm}
Consider the 2-body gadget Hamiltonian $\tilde{H}=H+V$ from \eqref{2localgadget_full},
corresponding to a target Hamiltonian $H_\text{targ}=H_\text{else}+\sum_{i=1}^M\gamma_iA_i\otimes B_i$ with $\gamma_j\le O(1)$ and $H_\text{else}$ positive semi-definite. 
Let $\Delta$ be the spectral gap between the ground and the first excited subspace of $H$,
and define a cutoff $\lambda_*=\Delta/2$.
Following Section \ref{sec:pert}, we define $\mathcal{L}_+=\text{span}\{\ket{\psi}:\bra{\psi}H\ket{\psi}>\lambda_*\}$ and $\tilde{\mathcal{L}}_-=\text{span}\{\ket{\psi}:\bra{\psi}\tilde{H}\ket{\psi}<\lambda_*\}$.
Then if $\Delta\ge 160M\gamma_\text{max}$, with $\gamma_\text{max}=\max_{j=1,\cdots,M}|\gamma_j|$, we have \[\tilde{\mathcal{L}}_-\cap\mathcal{L}_+=\varnothing.\] \end{claim}

To prove Claim \ref{Lpm}, it suffices to show that if $\ket{\psi}\in\mathcal{L}_+$ then $\bra{\psi}\tilde{H}\ket{\psi}>\lambda_*$ for any normalized state $\ket{\psi}$.
The subspace $\mathcal{L}_+$ is spanned by (direct + core) ancilla qubit states with at least one $\ket{1}$.
Let $\mathcal{K}_-=|0\cdots 0\rangle_w$ be the all-zero state the direct ancillas, and let $\mathcal{S}_a=\text{span}\{\ket{x}_\mathcal{C}:h(x)=a\}$ be the subspace of the core ancillas with exactly $a$ qubits\footnote{Here $h(x)$ is the Hamming weight of the binary string $x$.} in the state $|1\rangle$. Thus, the subspace $\mathcal{L}_+$ splits into two parts as $\mathcal{L}_+=\mathcal{L}_1\oplus\mathcal{L}_2$, where\footnote{Here $\mathcal{H}_w$ is the Hilbert space of the direct ancillas.}
\begin{align}
\label{eq:L12}
	\mathcal{L}_1 = \mathcal{H}_w \otimes \left(\bigoplus_{a=1}^C\mathcal{S}_a\right),\qquad\qquad
	\mathcal{L}_2 = \mathcal{K}_-^\perp \otimes\mathcal{S}_0.
\end{align}
The first part $\mathcal{L}_1$ spanned by all the states where the core has at least one qubit $|1\rangle$, while the second part $\mathcal{L}_2$ is spanned by all the states with the core ancillas all $|0\rangle$, and at least one direct ancillas being $\ket{1}$. We now first show that $\forall\ket{\psi}\in\mathcal{L}_1$, $\bra{\psi}\tilde{H}\ket{\psi}>\lambda_*$, and then similarly for $\mathcal{L}_2$.
$\quad$\\


\noindent{{\bf (1) }\emph{If $\ket{\psi}\in\mathcal{L}_1$, then $\bra{\psi}\tilde{H}\ket{\psi}>\lambda_*$.}}
$\quad$\\

\noindent Let us first consider $|\psi_a\rangle\in\mathcal{H}_w\otimes\mathcal{S}_a$ for some fixed $a\in\{1,2,\cdots,C\}$. Then $|\psi_a\rangle$ is a (linear combination of) state(s) where $a$ ancillas in the core are $|1\rangle$ and the other $C-a$ core ancillas are $|0\rangle$. We will find a lower bound for $\tilde{E}_{+,a}=\langle\psi_a|\tilde{H}|\psi_a\rangle$ by considering each component of $\tilde{H}$. Recall from \eqref{H2bodySingleCore} the definition of the core Hamiltonian
\begin{equation}\label{eq:Hcdef}
	H_\mathcal{C}=\frac{J}{2}\sum_{c\in \mathcal{C}}(\ii-Z_c) + \frac{J}{2}\sum_{c,c' \in \mathcal{C}}(\ii-Z_c Z_{c'}).
\end{equation}
Then the energy of $|\psi_a\rangle$ with respect to the core Hamiltonian is $E_{\mathcal{C},a}=\langle\psi_a|H_\mathcal{C}|\psi_a\rangle=Ja(C-a+1)\ge JC$. 
Let 
\begin{align}
    H_w = \frac{J}{2} \sum_{j=1}^{M}\sum_{i=1}^{R} \sum_{c\in \mathcal{C}}(\ii-Z_{w_i^{(j)}} Z_c)
     = \frac{J}{2} \sum_{j=1}^M\sum_{i=1}^R\left(C\ii-Z_{w_i^{(j)}}\sum_{c\in\mathcal{C}}Z_c\right)
     \label{directH}
\end{align}
be the interaction Hamiltonian between the direct ancillas and the core ancillas. Recall from \eqref{2localgadget_full} that $H=H_w+H_\mathcal{C}$. The second equality in \eqref{directH} indicates that $H_w$ consists of a sum of terms of the form $C\ii-Z_{w_i^{(j)}}\sum_{c\in\mathcal{C}}Z_c$. Let us focus on such a term for a particular direct ancilla $w$. Consider the states $\ket{0}_w \otimes \ket{a}_{\mathcal{C}}$ 
and $\ket{1}_w \otimes \ket{a}_{\mathcal{C}}$
with $\ket{a}_{\mathcal{C}} \in \mathcal{S}_a$ and look at the term 
$Z_w\sum_{c\in\mathcal{C}}Z_c$. Its expectation value in these states is $C-2a$ and $2a-C$, regardless of the state of the core ancillas.
Thus, we get an effective Hamiltonian 
\begin{equation}
  h'_w = C \ii -(C-2a)Z_w
\end{equation}
for each direct ancilla $w$.
Collecting these effective Hamiltonians for each direct ancilla, we get 
\begin{equation}\label{eq:Hpw}
H'_w=\frac{J}{2}\sum_{j=1}^M\sum_{i=1}^Rh'_{w_i^{(j)}}=\frac{J}{2}\sum_{k=1}^Nh'_k,
\end{equation}
whose lowest energy in the subspace $\mathcal{H}_w\otimes\mathcal{S}_a$ is equal to that of $H_w$. For convenience, we relabel the direct ancillas by $k = 1,\dots N$ with $N=MR$, and replace the sum over $i$ and $j$ with a single index summation over $k$.

Let us now add the perturbation $V$ \eqref{2localgadget_full}. For each direct ancilla $k$ there is a term in $V$ of the form $v_k=\beta_k(A\otimes X_k-B\otimes X_k)=\beta_k O_{AB}\otimes X_k$. Since we assume that $A$ and $B$ are unit-norm operators, the lowest energy of $v_k$ in $\mathcal{H}_k \otimes \mathcal{S}_a$ is lower bounded by that of $v'_k=2\beta_k X_k$. Thus, when we label $V'=\sum_{k=1}^N v'_k=\sum_{k=1}^N2\beta_k X_k$,
we get a 1-local Hamiltonian 
\begin{equation}\label{eq:1local}
\begin{array}{ccl}
	\tilde{H}' & = & \displaystyle E_{\mathcal{C},a}\ii + H'_w + V'
	= E_{\mathcal{C},a}\ii
	  + \sum_{k=1}^N\left(2\beta_kX_k+\frac{JC}{2}\ii-\frac{J(C-2a)}{2}Z_k\right) \\[0.1in]
& = & \displaystyle \left(Ja(C-a+1)+\frac{JCN}{2}\right)\ii
    + \sum_{k=1}^N\sqrt{4\beta_k^2+\frac{J^2}{4}(C-2a)^2} P_k
\end{array}
\end{equation} 
acting only on the direct ancillas, 
which gives us a lower bound on $E_+$, i.e. 
for any $|\psi_a\rangle\in\mathcal{H}_w\otimes\mathcal{S}_a$, 
\begin{equation}
	\min_{ \mathcal{H}_w\otimes\mathcal{S}_a}\langle\psi_a|\tilde{H}|\psi_a\rangle
	\ge
	\min_{ \mathcal{H}_w\otimes\mathcal{S}_a}\langle\psi_a|\tilde{H}'|\psi_a\rangle,
	\label{lowerboundEP}
\end{equation}
with $P_k$ a single qubit operator of the form $\hat{p}\cdot \vec{\hat{\sigma}}$, with $\vec{\hat{\sigma}} = \{X,Y,Z\}$ and unit vector $\hat{p}$. 
Note that the lower bound \eqref{lowerboundEP} does not include $H_{else}$ in $\tilde{H}'$; because $H_\text{else} \geq 0$,
we are only lowering the right side by omitting it. 

Note that the above argument can be generalized to $\mathcal{L}_1=\mathcal{H}_w\otimes(\oplus_{a=1}^C\mathcal{S}_a)$. For a general $\ket{\psi}\in\mathcal{L}_1$, $\ket{\psi}$ must take the form
\begin{equation}\label{eq:state_gen}
	\ket{\psi} = \ket{\phi}_w \otimes \sum_{a=1}^C\eta_a \ket{a}_\mathcal{C},
	\qquad\text{where}\qquad
	\ket{a} = \sum_{h(x)=a}c_{a,x} \ket{x},\quad x\in\{0,1\}^C
\end{equation}
for some sets of complex coefficients $\{\eta_a\}$ and $\{c_{a,x}\}$ that are both normalized. Then $\bra{\psi}H_\mathcal{C}\ket{\psi}=\sum_{a=1}^C|\eta_a|^2Ja(C-a+1)$. Let $\mathcal{A}$ be the set of $a$ for which $\eta_a\neq 0$. Let $a_\text{max}$ be the value of $a$ in $\mathcal{A}$ that maximizes $(C-2a)^2$. 
Define 
\begin{equation}
\begin{array}{ccl}
	|\psi'\rangle & = & \displaystyle \ket{\phi}_w\otimes|a_\text{max}\rangle_\mathcal{C}, \\[0.1in]
	h'_{w,a_\text{max}} & = & \displaystyle C\ii-(C-2a_\text{max})Z_w, \\[0.1in]
	H'_{w,a_\text{max}} & = & \displaystyle \frac{J}{2}\sum_{k=1}^Nh'_k.
\end{array}
\end{equation}
Then $\bra{\psi}H_w\ket{\psi}\ge\langle\psi_a|H'_{w,a_\text{max}}|\psi_a\rangle$ for any $\ket{\phi}_w\in\mathcal{H}_w$. Since the generalization from $\mathcal{H}_w\otimes\mathcal{S}_a$ to $\mathcal{H}_w\otimes(\oplus_{a=1}^C\mathcal{S}_a)$ does not concern the direct ancillas, we can use the same argument as before to construct a 1-local Hamiltonian
\begin{equation}
\tilde{H}'_{a_\text{max}}=\sum_{a=1}^C|\eta_a|^2\left(Ja(C-a+1)+\frac{JCN}{2}\right)\ii+\sum_{k=1}^N\sqrt{4\beta_k^2+\frac{J^2}{4}(C-2a_\text{max})^2}P_{k,a_\text{max}}
\label{eq:Hamax}
\end{equation}
such that for any $|\psi_a\rangle\in\mathcal{L}_1$, there always exists a value $a_\text{max}$ such that 
$\min_{\ket{\psi}\in \mathcal{L}_1}\bra{\psi}\tilde{H}\ket{\psi}\ge\min_{\ket{\psi}\in \mathcal{L}_1 }\langle \psi|\tilde{H}'_{a_\text{max}}\ket{\psi}$.

Let us now find a lower bound on $\langle \psi|\tilde{H}'_{a_\text{max}}\ket{\psi}$. Note that $Ja(C-a+1)\ge JC$ for any $a=1,2,\cdots,C$. Let $\beta_\text{max}=\max_{k=1,2,\cdots,N}|\beta_k|$. Noting that $P_{k,a_\text{max}}$ in \eqref{eq:Hamax} is a unit-norm operator, for any $\ket{\psi}\in\mathcal{L}_1$ we get
\begin{align}
	\bra{\psi}\tilde{H}'_{a_\text{max}}\ket{\psi} 
	& \ge  \displaystyle \left(JC+\frac{JCN}{2}\right) - N\sqrt{4\beta_\text{max}^2+\frac{J^2}{4}(C-2a_\text{max})^2} \nonumber\\
	& =   JC + \frac{JCN}{2}-\frac{JCN}{2}\sqrt{1+\frac{16\beta_\text{max}^2}{J^2C^2}} \nonumber\\
	& \ge   JC - \frac{JCN}{2}\cdot\frac{16\beta_\text{max}^2}{2J^2C^2}=JC-\frac{4N\beta_\text{max}^2}{JC} \nonumber\\
	& =   \Delta - \frac{4MR\beta_\text{max}^2}{\Delta}
	 \ge   \frac{79\Delta}{80} 
   	>\frac{\Delta}{2} = \lambda_*, \label{eq:L1bound2}
\end{align}
where we have used 
$2R\beta^2_\text{max}/\Delta = \gamma_\text{max}$ from \eqref{betachoice} and asked for $\Delta\ge 160M\gamma_\text{max}$ in the last line. 
Here $\gamma_\text{max}=\max_{j=1,\cdots,M}|\gamma_j|$ where $\gamma_j$ are coefficients in the target Hamiltonian. 
Putting \eqref{eq:L1bound2} 
into \eqref{lowerboundEP}, we get $E_+ > \frac{\Delta}{2} = \lambda_*$.
We have thus shown the desired lower bound on $E_+$ in the subspace $\mathcal{L}_1$. Let us now deal with the other part, $\mathcal{L}_2$. 

$\quad$\\
\noindent{{\bf (2) }\emph{If $\ket{\psi}\in\mathcal{L}_2$, then $\bra{\psi}\tilde{H}\ket{\psi}>\lambda_*$.}}
$\quad$\\

\noindent Any state in the subspace $\mathcal{L}_2 = \mathcal{K}_-^\perp \otimes\mathcal{S}_0$ has the core ancillas in the state $|0\cdots 0\rangle_\mathcal{C}$, hence $\bra{\psi}H_\mathcal{C}\ket{\psi}=0$. To find a lower bound for the energy of $H_w$ in this subspace, we use the construction $H'_w$ in \eqref{eq:Hpw} with $a=0$. For the energy of $V$ we use the same simplifying argument and obtain (again) a 1-local Hamiltonian acting only on the direct ancillas (cf.\ Equation \ref{eq:1local})
\begin{align}
	\tilde{H}'_0 = \sum_{k=1}^N\left(\frac{\Delta}{2}\ii-\frac{\Delta}{2}Z_k-2\beta_kX_k\right)=\sum_{k=1}^NS_k,
\label{eq:1local2}
\end{align}
such that
\begin{equation}\label{eq:1local2bound}
	\min_{\ket{\psi}\in\mathcal{L}_2}\bra{\psi}\tilde{H}\ket{\psi}
	\ge
	\min_{\ket{\psi}\in\mathcal{L}_2}\bra{\psi}\tilde{H}'_0\ket{\psi}.
\end{equation}
We now show that the energy of any direct ancilla state orthogonal to $\mathcal{K}_- = \text{span}\{|0\dots 0\rangle_w\}$ is strictly lower bounded by $\lambda_*=\Delta/2$. Since the core ancilla state will always be $|0\cdots 0\rangle_\mathcal{C}$, we will exclude it from our discussion and thus omit the $w$ subscript for the direct ancilla. All quantum states in the proof from here on refer to the state of the direct ancillas. 

To show the energy lower bound we use induction on the number of direct ancillas, $n$. Let 
\begin{equation}
E_n=\min_{\ket{\phi}\perp|0\rangle^{\otimes n}}\langle\phi|\sum_{k=1}^nS_k\ket{\phi}.
\end{equation}
Specifically, we prove the following statement:
\begin{align}
	E_n \geq \frac{3\Delta}{4}  - \delta_n, 
	\qquad \textrm{with} \qquad 
	\delta_{n} =  \frac{40n\beta^2}{9\Delta},
	\qquad 
	n=1,\cdots,N.  \label{ENbound1} 
\end{align}
We start with the initial case $n=1$. There the only state orthogonal to $\ket{0}$ is $\ket{1}$. Hence $E_1 = \Delta$, which satisfies \eqref{ENbound1}. 
Now assume \eqref{ENbound1} holds for some $n$. An $(n+1)$-qubit state that is orthogonal to $\ket{0\cdots 0}$ (denoted by the superscript $\text{\o}$) must have the form
\begin{align}
	\ket{\psi^{\text{\text{\o}}}_{n+1}} = a \ket{\xi^{\text{\text{\o}}}_{n}} \ket{0} + b \ket{\phi^{\text{\text{\o}}}_{n}} \ket{1} + c \ket{0\cdots 0}\ket{1}, \label{candidateNplus1}
\end{align}
where $\ket{\xi^{\text{\text{\o}}}_{n}}$ and $\ket{\phi^{\text{\text{\o}}}_{n}}$ are some states that are orthogonal to $\ket{0\cdots 0}$.
Let us calculate the energy of the state \eqref{candidateNplus1}.
\begin{align}
	E_{n+1} &= \sum_{i=1}^n \bra{\psi^{\o}_{n+1}} S_i \ket{\psi^{\o}_{n+1}} 
						+ \bra{\psi^{\o}_{n+1}} S_{n+1} \ket{\psi^{\o}_{n+1}}\\
	&= 
	    |a|^2 \sum_{i=1}^{n} \bra{\xi^{\text{\text{\o}}}_{n}} S_i \ket{\xi^{\text{\o}}_{n}} \braket{0}{0}
		+ |b|^2 \sum_{i=1}^{n} \bra{\phi^{\text{\text{\o}}}_{n}} S_i \ket{\phi^{\text{\o}}_{n}} \braket{1}{1} 
		+ |c|^2 \sum_{i=1}^{n} \bra{0\cdots 0}S_i \ket{0\cdots 0} \braket{0}{0}
	\\
	&	+ 2\textrm{Re}\left(
					ab^* \sum_{i=1}^{n}\bra{\xi^{\text{\text{\o}}}_{n}}S_i\ket{\phi^{\text{\o}}_{n}} \braket{0}{1}  
				+ ac^* \sum_{i=1}^{n}\bra{\xi^{\text{\o}}_{n}}S_i\ket{0\cdots 0} \braket{0}{1}  
				+ bc^* \sum_{i=1}^{n}\bra{\phi^{\text{\o}}_{n}}S_i\ket{0\cdots 0} \braket{1}{1} 
				\right) \nonumber\\
	& + |a|^2 \bra{0}S\ket{0}_{n+1} 
	  + |b|^2 \bra{1}S\ket{1}_{n+1} 
		+ |c|^2 \bra{1}S\ket{1}_{n+1} \nonumber\\
	&	+ 2\textrm{Re}\left(
				ab^* \braket{\xi^{\text{\o}}_n}{\phi^{\text{\o}}_n} \bra{0}S\ket{1}
				+ ac^* \braket{\xi^{\text{\o}}_n}{0\cdots 0} \bra{0}S\ket{1} 
				+ bc^* \braket{\phi^{\text{\o}}_n}{0\cdots 0} \bra{1}S\ket{1}
		\right). \nonumber 
\end{align}
Note that $\braket{0}{1}=0$ and $\braket{\psi_{n}^{\text{\o}}}{0\cdots 0} = \braket{\phi_{n}^{\text{\o}}}{0\cdots 0} = 0$.
Also recall that $\bra{0}S_i\ket{0}=0$, $\bra{1}S_i\ket{1}=\Delta$ and $\bra{0}S_i\ket{1}=-2\beta_i$. Hence,
\begin{align}
	E_{n+1} 
	& = 
	|a|^2 \sum_{i=1}^{n} \bra{\xi^{\text{\o}}_{n}} S_i \ket{\xi^{\text{\o}}_{n}} 
		+ |b|^2 \sum_{i=1}^{n} \bra{\phi^{\text{\o}}_{n}} S_i \ket{\phi^{\text{\o}}_{n}} + 0 
					+ 0 + 0
					+ 2\textrm{Re}\left( bc^* 
								\sum_{i=1}^{n}\bra{\phi^{\text{\o}}_{n}}S_i\ket{0\cdots 0} 
						\right) 
						 \\
	 & + 0 + |b|^2 \Delta + |c|^2 \Delta 
		+ 2\textrm{Re}\left(
				- ab^* \braket{\xi^{\text{\o}}_n}{\phi^{\text{\o}}_n} 2\beta_{n+1}
		\right) + 0 + 0 \nonumber \\
	&\geq |a|^2 E_n + |b|^2 E_n 
				+ 2\textrm{Re}\left( bc^* 
								\sum_{i=1}^{n}\bra{\phi^{\text{\o}}_{n}}S_i\ket{0\cdots 0} 
						\right) 
						 + |b|^2 \Delta + |c|^2 \Delta 
		- 4|a||b| \beta_\text{max}, \label{minimizeENfull} 
\end{align}
where we lower bounded the last term using absolute values, a maximum magnitude of the $\beta$'s, and $\left|\braket{\psi^{\text{\o}}_n}{\phi_n^{\text{\o}}}\right|\leq 1$. Next, we observe that the term $\bra{\phi^{\text{\o}}_{n}}S_i\ket{0\cdots 0}_{n}$ is nonzero only for parts of $\ket{\phi^{\text{\o}}_{n}}$ with a single $\ket{1}$. The largest magnitude it could possibly have is when the state $\ket{\phi^{\text{\o}}_{n}}$ is made {\em only} from states with a single $\ket{1}$ as $\frac{1}{\sqrt{n}} \sum_{i=1}^{n} \ket{0\cdots 1_i \cdots 0}$.
We then get $\sum_{i=1}^{n}\bra{\phi^{\text{\o}}_{n}}S_i\ket{0\cdots 0} \geq -2\beta_\text{max} \sqrt{n}$. Putting this in, recalling \eqref{ENbound1} and using absolute values, we get
\begin{align}
	E_{n+1} 
		& \geq \left( |a|^2 + |b|^2\right) \left( \frac{3\Delta}{4} - \delta_n \right) 
				- |b||c| \cdot 4\beta_\text{max} \sqrt{n}  
				+ \left(|b|^2 + |c|^2 \right) \Delta - |a||b| \cdot 4\beta_\text{max}  \\
	 &= \frac{3\Delta}{4} \left(|a|^2 + |b|^2 + |c|^2 \right) 
	   - \left(|a|^2+|b|^2\right) \delta_n
				+ |b|^2 \Delta
				- |a||b| \cdot 4\beta_\text{max} \nonumber
				- |b||c| \cdot 4\beta_\text{max} \sqrt{n}  
				+ |c|^2 \cdot \frac{\Delta}{4}
				 \\
	&\geq \frac{3\Delta}{4} - \delta_n 
			+ |b|^2 \Delta 
			- |b|\cdot 4\beta_\text{max}   
				+ \underbrace{|c| 
						\left( |c|\frac{\Delta}{4} - |b| \cdot 4\beta_\text{max}\sqrt{n} \right)}_{f(|c|)},
			\label{minimizeFULL3}
\end{align}
where we have used \eqref{ENbound1}, and then $|a|^2+|b|^2\leq 1$ and $|a|\leq 1$.
Independent of $|b|$, let us look at $f(|c|)$, a quadratic function of $|c|$. Its minimum is at $|c| = \frac{|b| \cdot 8\beta_\text{max} \sqrt{n}}{\Delta}$, with the value $f_\text{min} = -\frac{ |b|^2\cdot 16n \beta_\text{max}^2}{\Delta}$. In \eqref{minimizeFULL3} it means

\begin{align}
	E_{n+1} 
	&\geq \frac{3\Delta}{4} - \delta_n 
			+ |b|^2 \Delta - |b|\cdot 4\beta_\text{max}  -\frac{16|b|^2n \beta_\text{max}^2}{\Delta} \\
	&= \frac{3\Delta}{4} - \delta_n  
			+ |b|^2 \underbrace{\left( \Delta 
							- \frac{16n \beta_\text{max}^2}{\Delta}\right)}_{\geq 9\Delta / 10} - |b|\cdot 4\beta_\text{max} \\
	& \geq \frac{3\Delta}{4} - \delta_n 
			+ \underbrace{
			|b|\left( |b| \frac{9\Delta}{10} - 4\beta_\text{max} \right)
			}_{g(|b|)},
			\label{minimizeFULL4}
\end{align}
where in the second line we have used $\Delta \geq 160 M\gamma_\text{max}$ 
to guarantee
$\Delta - \frac{16\beta_\text{max}^2 n}{\Delta} \geq \Delta - \frac{16 \beta_\text{max}^2 N}{\Delta}
= \Delta - 16 M\gamma_\text{max} \geq \frac{9\Delta}{10}$. 
The expression $g(|b|)$ is quadratic in $|b|$, minimized at 
$|b| = \frac{20\beta_\text{max}}{9\Delta}$, giving the value
$g_\text{min} = -\frac{40\beta_\text{max}^2}{9\Delta}$. Putting it into \eqref{minimizeFULL4}, we get
\begin{align}
	E_{n+1} 
	&\geq  \frac{3\Delta}{4} - \delta_n -\frac{40\beta_\text{max}^2}{9\Delta} 
	= \frac{3\Delta}{4} - \delta_{n+1},  
\end{align}
which proves our induction step, as $\delta_n = \frac{40 n\beta_\text{max}^2}{9\Delta}$.
Therefore, \eqref{ENbound1} holds. Let $n=N$ and we have for any $\ket{\psi}\in\mathcal{L}_2$, 
\begin{align}
	\bra{\psi}\tilde{H}'_0\ket{\psi} \geq E_N &\geq \frac{3\Delta}{4} - \frac{40N\beta_\text{max}^2}{9\Delta} 
	 = \frac{3\Delta}{4} -  \frac{40MR\beta_\text{max}^2}{9\Delta} \nonumber \\
	 &= \frac{3\Delta}{4} -  \frac{20M\gamma_\text{max}}{9} 
	 \geq \left(\frac{3}{4}-  \frac{1}{72}\right) \Delta = \frac{53}{72} \Delta
	 > \frac{\Delta}{2} = \lambda_*,
    \label{eq:ep_bound}
\end{align}
where in the last line we have used \eqref{betachoice} and $\Delta\ge 160M\gamma_\text{max}$. Combining the above statement with \eqref{eq:1local2bound}, we have $\bra{\psi}\tilde{H}\ket{\psi}>\lambda_*$
for any $\ket{\psi}\in\mathcal{L}_2$. This concludes the proof of Claim \ref{Lpm}. 
\rightline{$\Box$}


\subsection{The perturbation series converges.}
\label{sec:perturbation2body}

Let us now state and prove our second claim -- the convergence of the perturbation series for our gadget construction.

\begin{claim}
\label{claim:perturb}
Consider the 2-body gadget Hamiltonian $\tilde{H}=H+V$ defined in \eqref{2localgadget_full} with spectral gap $\Delta$ between the ground and the first excited subspace of $H$, and a target Hamiltonian $H_\text{targ}=H_\text{else}+\sum_{j=1}^M\gamma_j A_j\otimes B_j$ with $\gamma_j=O(1)$ and $H_\text{else}$ positive semi-definite. Choose a constant parameter $d\in(0,1)$ and an error tolerance $\epsilon$. 
If we set $\Delta=M^3R^d$ and choose the number of direct ancillas per target term $R$ and the core size $C$ according to
\begin{align}
\label{eq:RC}
	R & \gg \max\left\{
			\epsilon^{-\frac{2}{d}}, 
			\left(\frac{\|H_\text{else}\|^2}{2M^4 \gamma_{\text{max}}}\right)^{\frac{1}{d}},
			\left(M^3 \epsilon^{-2} \right)^{\frac{1}{1-d}}
		\right\}, \qquad C\gg M^3 R^d \,\epsilon^{-1},
\end{align}
then the strengths of the interaction terms in the gadget Hamiltonian are small, i.e. $\beta_j, J =O(\epsilon)$.

Furthermore, the self energy expansion 
\eqref{eq:selfenergy} satisfies
\begin{equation}
	\|\Sigma_-(z)-H_\text{targ}\otimes\Pi_-\| = O(\epsilon),
\end{equation}
where $\Pi_-$ is the projector onto $\mathcal{L}_-$, and $z$ obeys $|z|  \leq \epsilon+\|H_\text{else}\|+\sum_{j=1}^M |\gamma_j|$.
\end{claim}
This claim is one of the central results of this work -- it shows that our gadget Hamiltonian (for a 2-local target Hamiltonian) uses  only interactions of strength $O(\epsilon)$, \emph{i.e.} no strong interactions.
This is qualitatively different from previous constructions which require interactions of strength poly$(\epsilon^{-1})$. However, the price we pay for avoiding strong interactions is that the number of ancillas scales as poly$(\epsilon^{-1})$, as shown in \eqref{eq:RC}, while previous constructions require some number of ancillas independent of $\epsilon$. Hence we present a tradeoff between interaction strength and ancilla number in a gadget Hamiltonian.

Let us prove Claim~\ref{claim:perturb}. First we show that $\tilde{H}$ consists of only weak interaction terms. 
When we choose $\Delta = M^3 R^d$ for some $d\in (0,1)$
and substitute it into \eqref{betachoice}, we find that the interaction strength 
 between the target spins and direct ancillas will be 
 $\beta_j = \sqrt{\frac{\gamma_j\Delta}{2R}} =O(\epsilon)$, if we choose 
 \begin{align}
 	\label{RconditionBeta2}
	R \gg \left(M^3 \epsilon^{-2} \right)^{\frac{1}{1-d}}.
\end{align}

Next, recalling $\Delta=CJ$, the strength of the interaction $J$ between the core ancillas will be $O(\epsilon)$ 
if we choose $C\gg M^3 R^d\,\epsilon^{-1}$. 

Furthermore, once we set $\Delta = M^3 R^d$, we can easily satisfy the requirement $\Delta \geq 160 M \gamma_{\max}$ in Claim~\ref{Lpm} for reasonable $R$ -- more specifically, we need $R\gg \left(160\gamma_\text{max}/M^2\right)^{1/d}$.

We will now analyze the higher order terms in the self energy expansion $\Sigma_-(z)$ according to \eqref{eq:selfenergy} and show that the error term in Equation \ref{eq:selfenergy} scales as $O(\epsilon)$. The perturbative expansion of $\Sigma_-(z)$ for the construction in \eqref{2localgadget_full} yields
\begin{align}
	\Sigma_-(z) = H_\text{else} + \frac{1}{z-\Delta} \sum_{j=1}^MR\beta_j^2(A_{a_j}-B_{b_j})^2
	 + \underbrace{\sum_{k=1}^\infty V_{-+} \left(G_+V_+\right)^k G_+ V_{+-}}_\text{error}.
	\label{eq:par_2nd}
\end{align}
We can associate every term  
in the perturbation series with a transition process starting in the ancilla state $\ket{0}_w\ket{0}_\mathcal{C}$ (i.e. belonging to $\mathcal{L}_-$) to states in $\mathcal{L}_+$ and back to $\mathcal{L}_-$. Specifically, at the $k$-th order, such transitions are 
\begin{align}\label{eq:legal_term}
	\mathcal{L}_- \xrightarrow{V_{-+}} \ket{y} 
	\underbrace{
		\xrightarrow{V_{+}}
		|y_1\rangle
		\xrightarrow{V_{+}}
		|y_2\rangle
		\xrightarrow{V_{+}}
		\cdots
		\xrightarrow{V_{+}}
		|y_{k-2}\rangle
		\xrightarrow{V_{+}}
	}_\text{$(k-2)$ steps}  
	\ket{y'} 
	\xrightarrow{V_{+-}}
	\mathcal{L}_-,
\end{align}
where both $y$ and $y'$ are $r$-bit strings with Hamming weight 1, and $|y_i\rangle\in\mathcal{L}_+$. Specifically, they belong to the subspace $\mathcal{L}_2= \mathcal{K}_-^\perp \otimes\mathcal{S}_0$ in \eqref{eq:L12}. In a transition, the $V_+$ component of the perturbation $V$ either flips a direct qubit $w$ or leaves it unchanged. In the former case, $V_+$ contributes terms that contain the interaction with $w$ via $\dots \otimes X_w$. In the latter case, $V_+$ contributes terms that contain interaction with $w$ via $\dots \otimes \ket{1}\bra{1}_w$. In general, out of the $k-2$ transitions, the number of flips $k_f$ cannot exceed $k$. Furthermore, it must be even for the transition to terminate in $\mathcal{L}_-$. Every transition step $|y_i\rangle\rightarrow|y_{i+1}\rangle$ also contributes a factor $\frac{1}{z-h(y_i)\Delta}$ to the corresponding term in the expansion $\Sigma_-(z)$, where $h(y_i)$ is the Hamming weight of the string $y_i$. 


\subsubsection{A simplified calculation without $H_\text{else}$}

\begin{figure}
\begin{center}
\includegraphics[width=12cm]{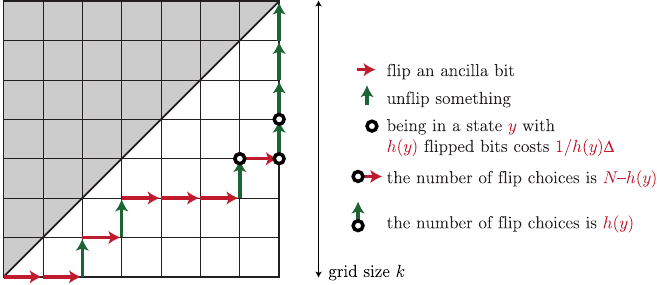}
\caption{Upper bounding the contributions to the error term at order $2k$ for a 2-body gadget without $H_{else}$.  
Each up- and right-moving path corresponds to a sequence of flips on the direct ancillas.
We are overcounting here, allowing paths that touch the diagonal, as well as paths that wander farther than $N$ from the diagonal when $k>N$.}
\label{fig:paths}
\end{center}
\end{figure}

To bound the norm of the error, let us start by considering the simplified case where $\|H_\text{else}\|=0$. The general calculation is presented below in Section~\ref{sec:generalPerturbationErrorBound}. 

In this simpler case, we have $\|V_+\|=0$ and all terms at odd orders vanish since $k_f$ must be even. Let $T_k=V_{-+}(G_+V_+)^{k-2}G_+V_{+-}$ be the $k$-th order term in $\Sigma_-(z)$. In Appendix~\ref{sec:pertorder246}, the reader can find the upper bound calculations for the first few terms of $\Sigma_-(z)$ as examples. Here, let us present the general even-order calculation.

Finding an upper bound for $\|T_{2k}\|$ for $k\geq 1$ amounts to enumerating all possible transitions and computing their respective contributions. For this, we introduce a graphical representation of the $(2k)^\text{th}$ order transitions in Figure \ref{fig:paths}. Each grid point in the lower-right triangle, including the diagonal points, corresponds to a state of the direct ancillas. We start from the lower-leftmost point, which corresponds to the all-zero subspace $\mathcal{L}_-$. Each step of the transition maps to a rightwards or upwards movement on the graph. A valid transition process ends at the top-rightmost point, which again belongs to the ground state subspace $\mathcal{L}_-$. Furthermore, a valid path can touch the diagonal line only at the last step of the transition. Suppose at a certain point the direct ancillas are in a state $|y\rangle$ with $h(y)$ ancillas in $|1\rangle$ and the rest in $|0\rangle$, with $h(y)\in\{1,2,\cdots,N\}$ being the Hamming weight of $y$. Then the next step will take $|y\rangle$ to a new state $|y'\rangle$ where 
$y$ and $y'$ differ by one bit. If 
an ancilla in $|y\rangle$ is flipped from $|0\rangle$ to $|1\rangle$ then we move to the right on the graph. If 
an ancilla is flipped from $|1\rangle$ to $|0\rangle$, we move up.

Since each \emph{fixed} path in the graph only records the change in the number of the direct ancillas that are in the state $|1\rangle$, our graphical representation is oblivious to the \emph{order} in which each of the direct ancillas are flipped. For a given path, we overcount the number of ways the ancillas were flipped ($|0\rangle\rightarrow|1\rangle$, horizontal lines in Figure \ref{fig:paths}), using $N^k$ -- the number of ways to draw $k$ out of $N$ objects \emph{with replacement}. Next, we bound the contribution from $G_+$ by $\|G_+\|\le\frac{1}{|h(y)\Delta-z|}\le\frac{1}{\Delta}$ where $h(y)$ is the Hamming weight of the state $|y\rangle$ in which a particular flip step starts. For each $|1\rangle\rightarrow|0\rangle$ flip (vertical line in Figure \ref{fig:paths}) that brings $|y\rangle$ to $|y'\rangle$, the resolvent $G_+$ contributes a factor $\frac{1}{z-h(y)\Delta}$. Furthermore, there are in total $h(y)$ different choices of which ancilla to unflip at this point, which gives us a factor $h(y)$ that ``cancels''\footnote{Here ``cancel" means that the product $\frac{h(y)}{|h(y)\Delta-z|}$ is $O(\Delta^{-1})$.} the factor $h(y)$ in $\frac{1}{z-h(y)\Delta}$ from the resolvent. Therefore, the contribution of a path in Figure~\ref{fig:paths} to $\|T_{2k}\|$ is at most
\begin{equation}
	c_\text{path} \le \frac{N^k(2\beta_\text{max})^{2k}}{\Delta^{2k-1}}.
\end{equation}

We can upper bound the number of valid paths such as the one shown in Figure \ref{fig:paths} by the $k^\text{th}$ Catalan number -- the number of up- \& right-moving paths between corners of a square that don't pass beyond the diagonal.
Besides all the valid paths that contribute non-trivially to $\|T_{2k}\|$,
the Catalan number $C_k$ also counts paths that are illegal for us -- those that touch the diagonal at places other than the start and finish points, and paths that possibly move farther than $N$ from the diagonal (for large $k$).
In this way, we get the upper bound\footnote{The Catalan number $C_k$ is trivially upper bounded by $4^k$, the number of possibilities for a $2k$-step process with 2 choices at each step.}
\begin{equation}\label{eq:2kbound_0}
	\|T_{2k}\| 
	\le C_k\cdot c_\text{path}
	\le 4^k\cdot \frac{N^k(2\beta_\text{max})^{2k}}{\Delta^{2k-1}}
	 = 8^k \Delta \left(\frac{N \cdot 2\beta_{\text{max}}^2}{\Delta}\right)^k \frac{1}{\Delta^k}
	 = \Delta \left( \frac{8 M \gamma_{\text{max}}}{\Delta} \right)^{k},
\end{equation}
using \eqref{betachoice}. Note that this is small for $k\geq 2$, as $\Delta \gg M$.
Thus, the error term $\|T_{2k}\|$ for a model without $H_\text{else}$ decays exponentially with $k$, and the series $\sum_{k=1}^\infty\|T_{2k}\|$ converges. Moreover, because we can choose $\Delta \gg M$, it can be upper bounded by a small number.


\subsubsection{A general calculation including $H_{\text{else}}$}
\label{sec:generalPerturbationErrorBound}

Now we turn to the general case where $\|H_\text{else}\|\neq 0$. In this scenario, each term in $T_{m}=V_{-+}(G_+V_+)^{m-2}G_+V_{+-}$ is composed from transitions of the following three types
\begin{enumerate}
	\item a $|0\rangle\rightarrow|1\rangle$ flip of some direct ancilla qubit $w$, 
	\item a $|1\rangle\rightarrow|0\rangle$ flip of some $w$, 
	\item the state of the ancillas staying the same. 
\end{enumerate}
In the first two cases, $V_+$ (also $V_{-+}$ or $V_{+-}$) 
contributes the term from $V$ that flips the direct ancilla $w$ via $\dots \otimes X_w$. In the third case, the ancilla state stays the same, and $V_+$ brings us the factor $H_\text{else}$. 

\begin{figure}
\begin{center}
\includegraphics[width=12cm]{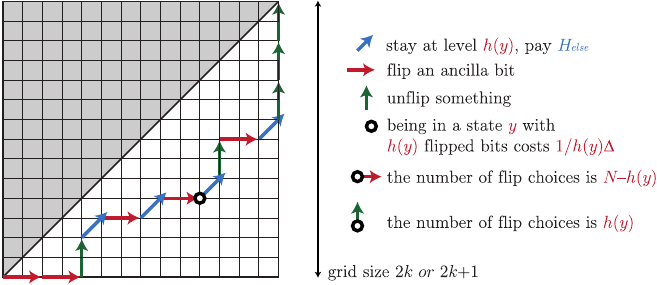}
\caption{Upper bounding the contributions to the 
error term of order $2k$ or $2k+1$. Up- and right-moving, subdiagonal paths
correspond to sequences of flips of direct ancillas. A bit flip moves 2 squares horizontally/vertically, ``staying'' moves across one square diagonally.} 
\label{fig:choose3}
\end{center}
\end{figure}

We modify our graphic approach from Figure \ref{fig:paths} to include the valid third type of transitions, corresponding to non-trivial terms in the expansion of $\Sigma_-(z)$. 
This approach is illustrated in Figure \ref{fig:choose3}. For a fixed $k$, the number of $2k$-step (resp.\ $(2k+1)$-step) transitions is now upper bounded by the \emph{Motzkin number} of order $2k$ (resp.\ $2k+1$).
These numbers correspond to a number of up-, diagonal-, and right-moving paths across a square, remaining below the diagonal.
Just as we easily upper bounded the Catalan numbers, it suffices for our purposes to upper bound the Motzkin numbers by $M_{2k}\le 3^{2k}$ and $M_{2k+1}\le 3\cdot 3^{2k}$.
\\


\noindent{\emph{Upper bounds on the $(2k)^\text{th}$ order.}} 
In estimating the error, we here consider only the $4^\text{th}$ order and onward, i.e. $k\ge 2$, as the second order is the actual term that we want to generate
 (for details of the 2nd order, see Appendix~\ref{sec:pertorder246}).

In order to make sure that the sequence of transitions finishes at $\mathcal{L}_-$, the number of flips $k_f=2f$ must be even ($f\in\mathbb{N}$, $f\le k$). Hence, the number of steps where the ancilla state stays the same is $2(k-f)$, an even number. A contribution from some path to $\Sigma_-(z)$ is a term whose norm is upper bounded by
\begin{align}
	\leq \frac{\overbrace{N(2\beta_\text{max})^2 \times \dots \times N(2\beta_\text{max})^2}^{f} \times \overbrace{\norm{H_\text{else}}^2\times \dots \times \norm{H_\text{else}}^2}^{k-f}}{\Delta^{2k-1}}.
	\label{boundpath1}
\end{align}
The condition 
$ R \ge  \left(\frac{\|H_\text{else}\|^2}{2M^4 \gamma_{\text{max}}}\right)^{\frac{1}{d}} $
 from \eqref{eq:RC} implies 
\begin{align}
	\|H_\text{else}\|^2 \le 2 M^4 R^d \gamma_{\text{max}} = 4 M R \frac{\Delta \gamma_{\text{max}}}{2R}  = N \left(2 \beta_{\text{max}}\right)^2.
	\label{RboundFromHelse}
\end{align}
 Therefore, the overall contribution of a path \eqref{boundpath1} is further upper bounded by
\begin{align}
	\leq \left(\frac{N(2\beta_\text{max})^2}{\Delta}\right)^k \frac{1}{\Delta^{k-1}}
	=2^k \left(M\gamma_{\text{max}}\right)^k \frac{1}{\Delta^{k-1}}
	= \Delta \left(\frac{2M \gamma_{\text{max}}}{\Delta}\right)^k.
\end{align}
The number of paths is upper bounded by $9^k$. Therefore, the overall norm of the $(2k)^\text{th}$ order is upper bounded by
\begin{align}
	\norm{T_{2k}} &\leq 
	9^k \Delta \left(\frac{2M\gamma_{\text{max}}}{\Delta}\right)^k
	= \Delta \left(\frac{18M\gamma_{\text{max}}}{\Delta}\right)^k.
	\label{eq:T2k}
\end{align}
We have chosen $\Delta\gg M$, so this is again a small contribution.
Note also the similarity between the form of the upper bound in the above result and \eqref{eq:2kbound_0}. 
\\


\noindent{\emph{Upper bounds on the $(2k+1)^\text{th}$ order.}} 
Bounding the $3^\text{rd}$ order is straightforward:
\begin{align}
	\|T_3\| = 
		N\cdot(2\beta_\text{max})\cdot\frac{1}{\Delta}\cdot\|H_\text{else}\|\cdot\frac{1}{\Delta}\cdot(2\beta_\text{max})
	\le \frac{\left(4 N \beta_{\text{max}}^2\right)^{\frac{3}{2}}}{\Delta^2}
	= \sqrt{\frac{\left(2M \gamma_{\text{max}}\right)^3}{\Delta}},
	\label{thirdorder}
\end{align}
using \eqref{betachoice}.
Recalling $\Delta = M^3 R^d$, we thus get $\|T_3\|\le(2\gamma_\text{max})^{3/2}R^{-d/2}=O\left(R^{-d/2}\right)$, a small contribution.

Similarly as above, we then calculate this for the $(2k+1)^\text{st}$ order in general and obtain
\begin{align}
 \|T_{2k+1}\| 
  & \le 3\cdot 9^k\cdot \frac{[N(2\beta_\text{max})^2]^f\cdot\|H_\text{else}\|^{2(k-f)+1}}{\Delta^{2k}} 
	\le 3\cdot 9^k\cdot\frac{[N(2\beta_\text{max})^2]^{k+\frac{1}{2}}}{\Delta^{2k}} \nonumber\\
  & \le 3\cdot 2^k \cdot 9^k \cdot \left(\frac{M\gamma_{\text{max}}}{\Delta}\right)^k 
  \sqrt{2 M\gamma_{\text{max}} \Delta}
  = \left(3\sqrt{2}\right) \Delta \left(
  	\frac{18M\gamma_{\text{max}}}{\Delta}\
  \right)^{k+\frac{1}{2}}
	\label{eq:T2kp1}
\end{align}

Comparing with \eqref{thirdorder}, we find that the last expression is also true for $k=1$. Therefore, using also \eqref{eq:T2k} we can bound all of the terms in the error series by 
\begin{align}
	\norm{T_m} &\leq 3\sqrt{2}\, \Delta \left(\frac{18M\gamma_{\text{max}}}{\Delta}\right)^{\frac{m}{2}}
	= 3\sqrt{2} \Delta q^m, 
\end{align}
for $m\geq 3$ with $q = \sqrt{18M\gamma_{\text{max}}/ \Delta} = O\left(M^{-1} R^{-\frac{d}{2}}\right)$.
Thus, the whole series $\sum_{m=3}^{\infty} \norm{T_m}$ is upper bounded by a geometric series that converges. We can upper bound it by 
\begin{align}
	\sum_{m=3}^{\infty} \norm{T_m} 
	\leq const. \times \Delta q^3 
	= O\left(R^{-\frac{d}{2}}\right)
	 \leq \epsilon,
\end{align}
for our choice of $\epsilon$ when we choose a suitably large $R \gg \epsilon^{-\frac{2}{d}}$. This concludes the proof of Claim~\ref{claim:perturb}.
\rightline{$\Box$}

In conclusion, in Equation \ref{eq:sigma_z} we have $\|\Sigma_-(z)-H_\text{eff}\|=O(\epsilon)$ 
where the effective Hamiltonian $H_\text{eff}=H_\text{targ}\otimes\Pi_-+\gamma\Pi_-$ (up to an overall shift) captures the target Hamiltonian.
Therefore we have proven Theorem~\ref{th2}. Let us have a last look at the resources we require for Theorem~\ref{th2} with the following remark.

\begin{remark}
If $H_\text{targ}$ acts on $n$ qubits, our gadget hamiltonian $\tilde{H}$ acts on 
\begin{align}
   n+MR+C & \gg  n+MR+M^3R^d\,\epsilon^{-1} \\
 	& \gg  n+\max\left\{
 	M\epsilon^{-\frac{2}{d}}+M^3\epsilon^{-3},\,
	\left( M^{4-d} \epsilon^{-2}\right)^{\frac{1}{1-d}}, \,
	M\left(\frac{\|H_\text{else}\|^2}{2M^4\gamma_\text{max}}\right)^{\frac{1}{d}}
	 + \epsilon^{-1}\,\frac{\|H_\text{else}\|^2}{2M\gamma_\text{max}}
	\right\} \nonumber
\end{align}
qubits. If the interaction graph of $H_\text{targ}$ has degree $D$, then the interaction graph of the gadget hamiltonian has total degree $\max\{DR,RC\}=\text{poly}(D,\epsilon^{-1},\|H_\text{else}\|,M)$.
\end{remark}

This concludes the story of the 2-body gadgets with weak interactions. Let us now apply the construction to reducing $k$-local to $2$-body with weak interaction $(k\ge 3)$.


\section{Reducing $k$-body to 2-body interactions $(k\ge 3)$}\label{sec:kbody}

\begin{figure}
\begin{center}
\includegraphics[width=13cm]{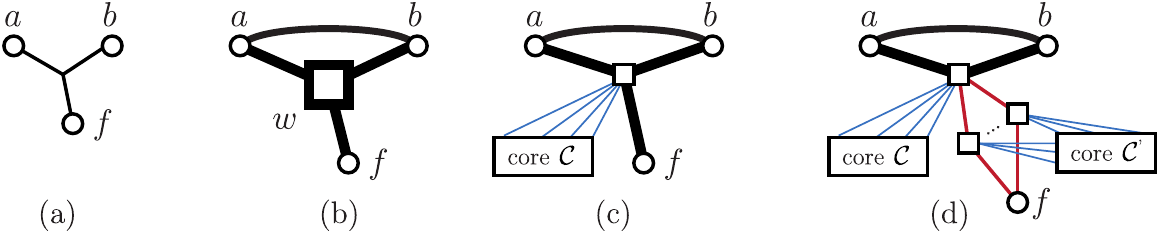}
\caption{3-local interactions from weak interactions. (a) The 3-local interaction we want to approximate. (b) The standard construction by Oliveira and Terhal \cite{OT06} with target term $A_{a}\otimes B_{b}\otimes F_{f}$ replaced by one (direct) ancilla $w$ in a large field $\Delta$, interacting with the target spins via strong interactions of order $\Delta^{2/3}$. In addition, $a$ and $b$ interact with strength of order $\Delta^{1/3}$ to compensate for the error term at $2^\text{nd}$ order perturbation theory. (c) The local fields are replaced by interactions with a core. (d) Each strong 2-local interaction term can be reduced to many $O(1)$ terms by our 2-body gadget construction, using another common core.
}
\label{fig:3body_sketch}
\end{center}
\end{figure}

With the new 2-body construction in mind, is it possible to use the \emph{core} idea and ``parallelism" of the 2-body gadgets to construct a 3-body to 2-body gadget that also uses only weak interactions? In Appendix \ref{sec:3body_problematic}, we show one such attempt. However, the resulting construction does not have all the desirable properties of the 2-body construction. We hope that the problems we have experienced will be useful for future efforts along similar lines. The difficulty to construct $k$-body gadgets along these lines is similarly difficult.

However, we are not doomed. Generating effective 3-body interactions such as $\gamma A\otimes B\otimes C$ using only weak interactions
is possible. We could simply apply our new 2-local construction on top of standard 3-body to 2-body gadgets as sketched in Figure \ref{fig:3body_sketch}. We start from the usual construction in \cite{OT06} and replace the strong 1-local term of magnitude $\Delta$ by interactions with a core. Finally, we reduce the large-norm 2-body interactions in these gadgets with weak ones using the 2-body gadgets from Section~\ref{sec:2body}.

For general $k$-body to 2-body reduction, we can resort to the construction from \cite{JF08}, where the gadget Hamiltonian consists of only 2-local interaction terms (\emph{i.e.\ }no extra 1-local terms). This makes it easy to directly apply our new 2-body gadgets and reduce the gadget Hamiltonian to one with only weak interactions. In subsequent work, we plan to elaborate on the detailed constructions and make them more efficient. 


\section{Acknowledgements}

We thank David Gosset, Dom Williamson, Ramis Movassagh, Sergei Bravyi and Sabre Kais for helpful discussions, and the Simons Institute for Theory of Computing at University of California, Berkeley, where the main part of this work was done. DN was supported by the Slovak Research and Development Agency grant APVV-0808-12 QIMABOS.
YC gratefully acknowledges support from Sabre Kais under the NSF Center for Quantum Information and Computation for Chemistry grant award number CHE-1037992.


\bibliographystyle{unsrt}
\bibliography{gadgetReferences}


\appendix


\section{Upper bounds on low-order perturbation series terms for 2-body gadgets}
\label{sec:pertorder246}

In this Appendix, for the purpose of illustration we calculate upper bounds on the norm of the first few orders in the perturbation series for the self-energy for our 2-body gadget construction from Section~\ref{sec:perturbation2body}.
\\

\noindent{\emph{The $2^\text{nd}$ order.}} 
This order is what contributes to the effective Hamiltonian, which has $M$ terms of norm $O(1)$ there. Let us see what we get here.
From \eqref{eq:par_2nd} we see that $T_2=\frac{1}{z-\Delta}\sum_{j=1}^MR\beta_j^2(A_{a_j}-B_{b_j})^2$. Every term at the second order corresponds to a transition of the form 
\begin{equation}\label{eq:2ndorder}
\mathcal{L}_-\rightarrow|y\rangle\rightarrow\mathcal{L}_-.
\end{equation}
Here $|y\rangle$ is a state where only one direct ancilla qubit $w$ is flipped to $|1\rangle$ while the others remain at $|0\rangle$. From our construction of $V$ in \eqref{2localgadget_full}, observe that each term that involves a particular direct ancilla $w_i^{(j)}$ is associated with a corresponding coefficient $\beta_j$. Therefore all the transitions of the form \eqref{eq:2ndorder} involving $w_i^{(j)}$ would contribute a term of the form
\begin{equation}\label{2ndordertrans}
\underbrace{\beta_j(A_{a_j}-B_{b_j})}_{V_{-+}}\cdot\underbrace{\frac{1}{z-\Delta}}_{G_+}\cdot\underbrace{\beta_j(A_{a_j}-B_{b_j})}_{V_{+-}}
\end{equation}
to the perturbative expansion $\Sigma_-(z)$. Note that because the Hamming weight of $y$ is $h(y)=1$, the resolvent component $G_+$ contributes a factor $\frac{1}{z-h(y)\Delta}=\frac{1}{z-\Delta}$. Since $R$ direct ancillas are introduced for the target 2-local term involving $a_j$ and $b_j$, the total contribution of the direct ancillas used for generating the $j$-th target term would be multiplied by a factor of $R$. Summing over all the target terms from $j=1$ to $M$, we get the current form of $T_2$. Assuming $A_{a_j}$ and $B_{b_j}$ are both unit-norm operators,
\begin{equation}
\|T_2\|\le\frac{1}{\Delta}\cdot MR(2\beta_\text{max})^2=2M\gamma_\text{max},
\end{equation}
using the choice $\beta_i=\sqrt{\frac{\gamma_i \Delta}{2R}}$. This is just what we expected (because the norm of what we are generating should be something on the order of $M$).

$\quad$\\
\noindent{\emph{The $4^\text{th}$ order.}} Transitions at the $4^\text{th}$ order could involve one or two direct ancillas\footnote{See also \cite{CRBK14} for a detailed explanation.}. In the former case the transition would take the form of
\begin{equation}
\mathcal{L}_-\rightarrow|y\rangle\rightarrow|y\rangle\rightarrow|y\rangle\rightarrow\mathcal{L}_-
\end{equation}
where $y$ is a string of Hamming weight 1. Such processes all contribute 0 to the perturbative expansion since $\|H_\text{else}\|=0$. 
Now we consider processes that involve two different direct ancillas $w_a$ and $w_b$. There are two possibilities:
\begin{align}
	\uparrow_a \uparrow_b \downarrow_a \downarrow_b, \qquad 
	\uparrow_a \uparrow_b \downarrow_b \downarrow_a
\end{align}
where $\uparrow_a$ means flipping $w_a$ from $|0\rangle$ to $|1\rangle$ and $\downarrow_a$ from $|1\rangle$ to $|0\rangle$. Similar for $w_b$. From $N=MR$ direct ancillas, there are in total $N(N-1)$ ways to choose $w_a$ and $w_b$. For a fixed choice of $w_a$ and $w_b$, each of the possible transitions listed above gives rise to at most $(2\beta_\text{max})^4$ from the 4 flipping processes (from the above discussion on \eqref{2ndordertrans} each flipping process contributes a factor of $2\beta_j\le 2\beta_\text{max}$ in $\|T_k\|$). The $G_+$ terms contribute an overall factor of $\frac{1}{z-\Delta}\cdot\frac{1}{z-2\Delta}\cdot\frac{1}{z-\Delta}$ to the perturbative expansion. In particular the factor 2 in the component $\frac{1}{z-2\Delta}$ is due to the fact that after the second flipping process the state has two ancillas flipped to 1, resulting in a state $|y'\rangle$ with $h(y')=2$. Combining these arguments, we have
\begin{equation}
\begin{array}{ccl}
\|T_4\| & \le & \displaystyle 2N(N-1)\cdot(2\beta_\text{max})^4\cdot\frac{1}{\Delta\cdot(2\Delta)\cdot\Delta}=\frac{N(N-1)(2\beta_\text{max})^4}{\Delta^3} \\[0.1in]
& \le & \displaystyle \left(\frac{N(2\beta_\text{max})^2}{\Delta}\right)^2\frac{1}{\Delta}=2M\gamma_\text{max}\cdot\left(\frac{2M\gamma_\text{max}}{\Delta}\right).
\end{array}
\end{equation}
Note that compared with the $2^\text{nd}$ order term, we collect a factor of $2M\gamma_\text{max}/\Delta$ in the upper bound for $\|T_4\|$.
$\quad$\\
$\quad$\\
\noindent{\emph{$6^\text{th}$ order.}} Following the same notation as before, at $6^\text{th}$ order the following transitions contribute non-trivially to $\|T_6\|$:
\begin{align}
	\uparrow_a \uparrow_b \uparrow_c (\downarrow)^3, \qquad 
	\uparrow_a \uparrow_b \downarrow_a \uparrow_c (\downarrow)^2.
\end{align}
The former type of transitions has $N(N-1)(N-2)\cdot 6$ different ways of occuring and the $G_+$ terms contribute a factor of $\frac{1}{z-\Delta}\cdot\frac{1}{z-2\Delta}\cdot\frac{1}{z-3\Delta}\cdot\frac{1}{z-2\Delta}\cdot\frac{1}{z-\Delta}$. The latter has $N(N-1)\cdot 2\cdot (N-1)\cdot 2$ different ways of occuring and a factor $\frac{1}{z-\Delta}\cdot\frac{1}{z-2\Delta}\cdot\frac{1}{z-\Delta}\cdot\frac{1}{z-2\Delta}\cdot\frac{1}{z-\Delta}$ from the $G_+$ components. Both types involve 6 flipping processes, which amounts to a factor of $(2\beta_\text{max})^6$. Hence
\begin{align}
	\norm{T_6} &= 6N(N-1)(N-2) (2\beta_\text{max})^6 \frac{1}{\Delta^2 (2\Delta)^2 (3\Delta)} + 4N(N-1)(N-1) (2\beta_\text{max})^6 \frac{1}{\Delta^3 (2\Delta)^2} \\
	&\leq \frac{3}{2} \left(\frac{N(2\beta_\text{max})^2}{\Delta}\right)^3 \frac{1}{\Delta^2}
	= 3M\gamma_\text{max} \left(\frac{2M\gamma_\text{max}}{\Delta}\right)^2. 
\end{align}
Note that another $2M\gamma_\text{max}/\Delta$ factor is collected at the $6^\text{th}$ order compared with the $4^\text{th}$. Given our choice that $\Delta=M^3R^d$, it is clear that $2M\gamma_\text{max}/\Delta=O(M^{-2}R^{-d})$. It is reasonable to speculate that $\|T_{2m}\|=O(M^{-2m}R^{-dm})$ converges exponentially as $m\rightarrow\infty$, which implies that the series $\sum_{m=2}^\infty\|T_{2m}\|$ converges.


\section{Justification for the sequence of simplifying Hamiltonians when bounding $E_+$}
\label{sec:EplusSimplification}

Let $\ket{\psi} \in \mathcal{L}_+$ be the state with minimum energy, i.e. $\bra{\psi} \tilde{H} \ket{\psi} = E_+$.
The Hamiltonian $\tilde{H}$ connects target spins to direct ancillas via terms of the type $A_a \otimes X_j$.
We now argue that $E_+$ can be only lowered if we 
decouple the target spins from the direct ancillas, 
and simply use $-\ii \otimes X_j$ instead of $A_a \otimes X_j$.

The expectation value $\bra{\psi} \tilde{H} \ket{\psi} = E_{H} + E_{V}$ comes from the expectation value of $H$ which is diagonal in the computational basis (the $Z$ and $ZZ$ terms involving the ancillas) and the expectation of $V$, which includes the interactions with target spins as well as the term $H_{\text{else}}$. 
Let us rewrite the state $\ket{\psi}$ as
\begin{align}
	\ket{\psi} = \sum_{w} c_w \ket{w} \otimes \ket{\phi_w},
\end{align}
where $w$ is a binary string labeling computational basis state of all the ancillas.
The expectation value of the term $H$ depends only on the magnitudes of the $c_w$'s. 
The contribution from $H_{\text{else}}$ is 
\begin{align}
	\sum_w \left| c_w \right|^2 \bra{\phi_w} H_{\text{else}} \ket{\phi_w}. \label{HELSEcontribute}
\end{align}
Finally, each term in $V$ of the form $A_a \otimes X_j$ contributes 
\begin{align}
	c_v^* c_{v'} \bra{v} X_j \ket{v'} \bra{\phi_v} A_a \ket{\phi_{v'}} \label{AXcontribute}
\end{align}
for every pair $v,v'$ that differ only at bit $j$. 
This expression can be positive or negative, depending on $c_v$ and $c_{v'}$.
More crucially, its magnitude will depend on $\bra{\phi_v} A_a \ket{\phi_{v'}}$. Because $A_a$ is a Pauli operator, this magnitude can never exceed 1. Let us now consider a state 
\begin{align}
	\ket{\psi'} = \left(\sum_{w} \left| c_w \right| \ket{w} \right) \otimes \ket{\phi},
\end{align}
with positive coefficients $|c_w|$, and a particular state $\ket{\phi}$ chosen to minimize $\bra{\phi} H_{\text{else}} \ket{\phi}$. The expectation value of $H$ does not change, while the contribution from $H_{\text{else}}$ can only decrease, because we have chosen $\ket{\phi}$ to minimize it. In other words, $\langle\psi'|H\psi'\rangle\le\langle\psi|H|\psi\rangle$ and $\langle\psi'|H_\text{else}|\psi'\rangle\le\langle\psi|H_\text{else}|\psi\rangle$. Finally, the expectation value of the interaction terms in $V'$ (when we set $A_a = -\ii$) like \eqref{AXcontribute} now become
\begin{align}
-|c_v|\cdot | c_{v'}| \bra{v} X_u \ket{v'} \le -c_v^*c_{v'}\bra{v} X_u \ket{v'} \langle{\phi_v}|A_a\ket{\phi_{v'}}. \label{AXcontribute2}
\end{align}

Thus, $\langle\psi'|V'|\psi'\rangle\le\langle\psi|V|\psi\rangle$ and we conclude that the new minimum energy of $\tilde{H'}$ restricted to $\mathcal{L}_+$ is $E'_+ \le  \bra{\psi'}H\ket{\psi'} + \bra{\psi'}V'\ket{\psi'} \leq  E_+$.
It means that when we replace the Hamiltonian $V$ with one that has no interactions between the direct ancillas and the target spins, and uses operators $-X_w$ on the direct ancillas, $E_+$ decreases (or remains what it was).


\section{An alternative idea for 3-local gadget construction}\label{sec:3body_problematic}

\begin{figure}
\begin{center}
\includegraphics[width=7cm]{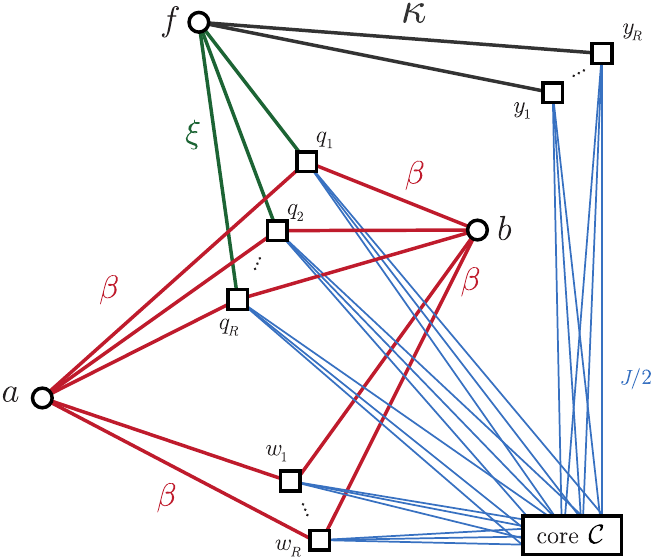}
\caption{A proposed effective three-body interaction on target spins $a,b,f,$ made from many weak 2-body interactions.
The construction \eqref{threebodyperturbation} involves a core of size $C$ and a set of direct ancillas $q_1,\dots,q_R$ whose 2-body interactions with the target spins mediate the desired 3-local interaction (but also some extra terms).
We then add two more sets of direct ancillas $w_1,\dots,w_R$ and $y_1,\dots,y_R$, in order to cancel out unwanted 2-local and 1-body terms. We can choose the interaction strengths $\beta, \xi, \kappa, J \ll 1$.} 
\label{fig:make3from2}
\end{center}
\end{figure}

Inspired by the 2-body gadget construction, one might naturally ask how to construct a similar gadget Hamiltonian for 3-local interactions. Consider a target 3-local term in a Hamiltonian $H_\text{targ}=H_\text{else}+\gamma A_a\otimes B_b\otimes F_f$
on three qubits $a,b,f$, with $\gamma=O(1)$ and $\|H_\text{else}\|=O(1)$.
Can we simulate this 3-local term by 2-local terms with small norms directly, without using an intermediate 3-body to strong-2-body gadget step? It turns out that we soon run into difficulties; we highlight them now and hope that they will be instructive for further work.

Let us try combining the 3-body to 2-body construction of Oliveira and Terhal \cite{OT06}
with the {\em core} and {\em parallelization} ideas from Section \ref{sec:2body}.
The resulting gadget (for a single 3-local term) is illustrated in Figure \ref{fig:make3from2}. Besides the three target particles, it contains a {\em core} $\mathcal{C}$ of $C$ qubits in a local field $\frac{J}{2}$, interacting with each other (as a complete graph) ferromagnetically with strength $\frac{J}{2}$. Next, we introduce three sets of (direct) ancilla qubits $q_i, w_i$ and $y_i$ (each set has size $R$), connect them to the core, and also let them interact with the three target spins $a,b,f$. 

Suppose we build a 2-body gadget Hamiltonian that from two parts, $\tilde{H}=H+V$, and treat $V$ as a perturbation.
We will generate the desired target interaction from the third order of the perturbation expansion.
Note that since no third order transition that starts from the ground state subspace of the direct ancilla and ends in it could involve more than one direct ancilla, the gadgets can be applied in parallel on $M$ target terms. Similar to \eqref{2localgadget_full}, we can then write the gadget construction for $M$ different 3-local target terms. 

To effectively capture a target 3-local Hamiltonian $H_\text{targ}=H_\text{else}+\sum_{j=1}^M\gamma_jA_{a_j}\otimes B_{b_j}\otimes F_{f_j}$, we use
\begin{align}
H & = \displaystyle \frac{J}{2}\sum_{j=1}^M\sum_{i=1}^R\sum_{c\in\mathcal{C}}(3\ii-Z_{w_i^{(j)}}Z_c-Z_{q_i^{(j)}}Z_c-Z_{y_i^{(j)}}Z_c) + H_\mathcal{C} \label{threebodynonperturb} \\
V & =  H_\text{else}+\sum_{j=1}^M\sum_{i=1}^R\Big( 
		\beta_j \left(A_{a_j}+B_{b_j}\right)\otimes X_{q_i^{(j)}}
	+ \xi_j F_{f_j} \otimes \ket{1}\bra{1}_{q_i^{(j)}} \label{threebodyperturbation} \\ 
  & \makebox[1.1in]{} +  \beta_j \left( A_{a_j} - B_{b_j} \right) \otimes X_{w_i^{(j)}}  
	+  \kappa_j \left(F_{f_j}-\ii\right) \otimes X_{y_i^{(j)}}
  \Big).	
	\nonumber
\end{align}
where $H_\mathcal{C}$ is the core Hamiltonian from \eqref{H2bodySingleCore}.
Working out the self energy expansion \eqref{eq:selfenergy}, we find 
\begin{equation}\label{eq:expand3body}
\begin{array}{ccl}
\Sigma_-(z) & = & \displaystyle H_\text{else}+ \underbrace{\frac{R}{z-\Delta}\sum_{j=1}^M\beta_j^2(A_{a_j}+B_{b_j})^2}_\text{(a)}
+ \underbrace{\frac{R}{z-\Delta}\sum_{j=1}^M\kappa_j^2(F_{f_j}-\ii)^2}_\text{(b)} \\
& + & \underbrace{\frac{R}{z-\Delta}\sum_{j=1}^M\beta_j^2(A_{a_j}-B_{b_j})^2}_\text{(c)}
+ \underbrace{\frac{R}{(z-\Delta)^2}\sum_{j=1}^M\beta_j^2\xi_j(A_{a_j}+B_{b_j})F_{f_j}(A_{a_j}+B_{b_j})}_\text{(d)} +\cdots
\end{array}
\end{equation}
Observe that the desired three-body interaction $\sum_{j=1}^M\frac{2R\beta_j^2\xi_j}{\Delta^2} A_{a_j}\otimes B_{b_j} \otimes F_{f_j}$ appears at the third order (d); it comes from the interplay of the terms involving $q_i$ in \eqref{threebodyperturbation}.
To make the 3-local coupling coefficient of the $j^\text{th}$ target term to be $\gamma_j$, we have to choose $\beta_j, \xi_j, \Delta,$ and $R$ (recalling that $\Delta = CJ$) so that 
\begin{align}
	\frac{2R \beta_j^2 \xi_j}{\Delta^2} = \gamma_j. \label{3bodySettings}
\end{align}

Besides our target 3-local interaction, the term (d) in the expansion of $\Sigma_{-}(z)$ \eqref{eq:expand3body} also gives us undesired interactions of the type $A_{a_j}\otimes B_{b_j}$ and a 1-local term $F_{f_j}$. 
Nevertheless, we can cancel them out -- the last two terms in \eqref{threebodyperturbation} compensate for these undesired terms by generating the terms (b) and (c) in \eqref{eq:expand3body}. The cancellation happens for the choice
\footnote{
Using the same prefactor $\beta$ in the first and third terms in \eqref{threebodyperturbation}
takes care of the term $A\otimes B$. 
Next, we have $\frac{R\beta^2\xi}{\Delta^2}\, F_f$ generated at the third order from the first two terms in \eqref{threebodyperturbation}. We cancel it out by $\frac{-2R\kappa}{\Delta} F_f$ generated at the second order from the last term in \eqref{threebodyperturbation}.} 
\begin{align}\label{betachoice_3body}
	\kappa_j = \sqrt{\frac{\beta_j^2 \xi_j}{\Delta}} = \sqrt{\frac{\gamma_j \Delta}{2R}},\qquad\qquad
  \xi_j = \beta_j = \left( \frac{\gamma_j \Delta^2}{2R} \right)^{\frac{1}{3}}.
\end{align}
This way, we obtain only the desired 3-local interaction (and $H_{\text{else}}$), and an overall energy shift.

Showing that this 3-local gadget satisfies the conditions of Theorem \ref{th:perturbation}
is now more difficult than what we did for the 2-body gadget.
It is not clear that the proofs (the subspace condition and convergence of the perturbation series) 
will go through with the interaction strengths $\xi_j, \beta_j, \kappa_j, J = O(\epsilon)$ for an input error parameter $\epsilon$.
However, it is possible that this construction will result in a better scaling of the necessary interaction strengths.
We leave this as an open question, and highlight the following difficulty.

Instead of \eqref{betachoice} for the 2-body gadget, in \eqref{betachoice_3body} we have a relationship between $\beta_j$ and $\Delta^2/R$, instead of $\Delta/R$. This difference has non-trivial consequences. If we were to proceed to prove that the 3-local gadget construction follows the subspace condition of the Theorem \ref{th:perturbation} as in Claim \ref{Lpm}, for subspaces $\mathcal{L}_1$ and $\mathcal{L}_2$ (see Equation \ref{eq:L12}) we could find the lower bound for $\langle\psi|\tilde{H}|\psi\rangle$ by constructing 1-local Hamiltonians whose energy is lower than $\langle\psi|\tilde{H}|\psi\rangle$ for any $|\psi\rangle\in\mathcal{L}_+$. Then for $\mathcal{L}_1$ we would obtain an inequality in the form of (in comparison with \eqref{eq:L1bound2})
\begin{equation}
\Delta > C\cdot N\beta_\text{max}
\end{equation}
for some constant $C$. Using the definition $\beta_\text{max}(\gamma_\text{max}\Delta^2/(2R))^{1/3}$, we have $\Delta = \Omega\left(M^3R^2\right)$. For $\mathcal{L}_2$ we would similarly obtain an inequality of the form (compare this to \eqref{eq:ep_bound})
\begin{equation}
\Delta > C'\cdot\frac{N\beta_\text{max}^2}{\Delta}
\end{equation}
for some constant $C'$. By the definition of $\beta_\text{max}$ we have $\Delta=\Omega\left(M^{3/2}R^{1/2}\right)$. Hence combining the asymptotic requirement for $\Delta$, we need $\Delta = \Omega\left(M^3R^2\right)$ for the gadget to satisfy the subspace condition of Theorem \ref{th:perturbation}. Bounding the terms in the perturbation expansion will also show that this asymptotic requirement for $\Delta$ is enough for the series to converge. However, in order for the coupling coefficients $J$, $\beta_i$, $\xi_i$ and $\kappa_i$ to be $O(\epsilon)$ for any $\epsilon>0$, we need each of them to scale as $O\left(R^d\right)$ where the exponent $d<0$. This could not be the case if we impose the above requirement for $\Delta$. For example, consider the $\beta_i$ coefficients. If we require that $\Delta=\Omega\left(M^3R^2\right)$,
\begin{equation}
	\beta_i = \left(\frac{\gamma_i\Delta^2}{2R}\right)^{\frac{1}{3}}
 = \Theta\left(R^{-\frac{1}{3}}\Delta^{\frac{2}{3}}\right)=\Omega\left(M^2R\right).
\end{equation}
This is where the gadget construction cannot consist of only weak terms. 
We hope that a way to circumvent this problem can be found in the future.


\subsection*{Reducing {\em k}-body to 2-body interactions}

Similarly, one can use standard {\em k}-body to 2-body gadgets to get a $2$-body Hamiltonian 
with large interaction strengths. After this, we apply the strong-from-weak gadget construction of Section~\ref{sec:2body}
and end up with $k$-local from weak $2$-body interactions. However, the efficiency is nothing to sing about.
Therefore, we propose to look at a direct, parallelized construction similar to Figure~\ref{fig:make3from2}, including interactions with a core to replace strong couplings. However, as noted above for the proposed $3$-body gadget, it will not be easy to show that the subspace conditions are met, the perturbation series converges, and the interactions remain weak.

\end{document}